\documentclass[useAMS,usenatbib,usegraphicx,dvipdfmx]{aa}

\usepackage{color}
\usepackage{hhline}
\usepackage{ulem}
\usepackage{comment}
\usepackage{amsmath}
\usepackage{lscape}

\def\mbf#1{\mbox{\boldmath ${#1}$}}

\def\lesssim{\; \underset{\sim}{<}\;}
\def\gtrsim{\; \underset{\sim}{>}\;}

\newsavebox\myVerb

\begin{document}
\title{Evolution of protoplanetary discs \\with magnetically driven disc winds}
\subtitle{}


\author{Takeru K. Suzuki
  \inst{1,2}
  \and
  Masahiro Ogihara
  \inst{3,4}
  \and
  Alessandro Morbidelli
  \inst{3}
  \and
  Aur\'{e}lien Crida
  \inst{3,5}
  \and
  Tristan Guillot
  \inst{3}
}
\institute{School of Arts \& Sciences, University of Tokyo,
  3-8-1, Komaba, Meguro, Tokyo 153-8902, Japan\\
  \email{stakeru@ea.c.u-tokyo.ac.jp}
  \and
  Department of Physics, Nagoya University, Nagoya, Aichi, 464-8602, Japan
  \and
  Laboratoire Lagrange, Universit\'{e} C\^{o}te d'Azur,
  Observatoire de la C\^{o}te d'Azur, CNRS, 
  Bvd de l'Observatoire, CS 34229, 06304 Nice Cedex 4, France
  \and
  Division of Theoretical Astronomy, National Astronomical
  Observatory of Japan, 2-21-1, Osawa, Mitaka, Tokyo 181-8588, Japan
  \and
  Institut Universitaire de France, 103 bd Saint Michel,
  75005 Paris, France
}
  
\titlerunning{Evolution of protoplanetary discs with magnetically driven disc winds}
\authorrunning{Suzuki et al.}




\abstract
{}
{We investigate the evolution of protoplanetary discs (PPDs hereafter)
  with magnetically driven disc winds and viscous heating.}
{
  We considered an initially massive disc with $\sim 0.1 M_{\odot}$ to track
  the evolution from the early stage of PPDs.
  We solved the time evolution of surface density and temperature by taking into
  account viscous heating and the loss of mass and angular momentum
  by the disc winds within the framework of a standard $\alpha$ model for
  accretion discs. 
  Our model parameters, turbulent viscosity, disc wind mass-loss, and disc
  wind torque, which were adopted from local magnetohydrodynamical simulations
  and constrained by the global energetics of the gravitational accretion,
  largely depends on the physical condition of PPDs, 
  particularly on the evolution of the vertical magnetic flux
  in weakly ionized PPDs.}
{
  Although there are still uncertainties concerning the evolution of
    the vertical magnetic flux that remains, 
  the surface densities show a large variety, depending on the combination
  of these three parameters, some of which are very different
  from the surface density expected from the standard accretion.
  When a PPD is in a wind-driven accretion state with the preserved
  vertical magnetic field, the radial dependence of the surface
  density can be positive in the inner region $<1-10$ au.
  The mass accretion rates are consistent with observations, even in
  the very low level of magnetohydrodynamical turbulence. 
  Such a positive radial slope
  of the surface density strongly affects 
  planet formation because 
  it inhibits the inward drift or even causes 
  the outward drift of pebble- to boulder-sized solid bodies, and it also slows
  down or even reversed the inward type-I migration of protoplanets.
   }
   {The variety of our calculated PPDs should yield a wide variety
  of exoplanet systems.}

   \maketitle
\section{Introduction}
The evolution of protoplanetary disks (PPDs) is one of the keys
to understand planet formation. There are still several unsolved problems,
one of which is the dispersal of PPDs
\citep[][]{hai01,her08,tak14,tak15}.
The evolution and dispersal of PPDs have been extensively studied in the
framework of viscously accreting discs that undergo 
photoevaporation by the irradiation
from the central star \citep[e.g,][]{shu93,hol00,alx06,kim16}.

In addition to the viscous accretion and the photoevaporation, the role of
magnetically driven disc winds has recently been received new attention.
\citet{si09} and \citet{suz10} proposed that vertical outflows driven by
magnetohydrodynamical (MHD; hereafter) turbulence might be a viable mechanism
that disperses the gas component of PPDs; turbulence is triggered by
magnetorotational instability \citep[MRI hereafter; ][]{vel59,cha61,bh91},
and the Poynting flux associated with the MHD turbulence drives vertical
outflows. The idea of MHD turbulence-driven outflow has also been 
extended by considering various effects, such as 
a stronger magnetic field \citep{bs13a}, a large-scale magnetic field
  \citep{les13}, and the dynamics of dust grains \citep{miy16},
whereas its mass flux is still quantitatively uncertain
\citep{fro13}.

Although \citet{suz10} considered mass loss to be the sole role of the disc
wind, the disc wind in reality also carries off the angular
momentum \citep{bp82,pp92,fer06,sal11}. In particular, a dead zone, which
is an MRI-inactive
region because of the insufficient ionization, is supposed to form in a
PPD \citep{gam96,san00}. In a dead zone the level of the excited turbulence
is low, and it is not sufficient to sustain the observed mass accretion onto
the central star.
In these circumstances, the extraction of the angular momentum by the disc wind
possibly plays a primary role in driving mass accretion \citep{bs13b,sim13}.
\citet{bai16a} and \citet{bai16b} investigated the global evolution of PPDs
in such a wind-driven accretion state, by also taking the effect of external
heating into account, and reported that a large portion of the mass
is removed by the disc wind in comparison to the accreting mass.

A critical open question concerning the disc wind from PPDs is that
the mass-loss rate.
At the later stage of the evolution, a wind footpoint that is determined by
the irradiation from a central star is expected to primarily control the
mass-loss rate \citep{bai16a,bai16b}.
On the other hand, at the earlier stage when the surface density is high,
viscous heating plays an essential role in determining the thermal properties of
PPDs \citep[e.g.,][]{rl86,nn94,ht11,oka11,bit15}. 
To investigate the time evolution from the early epoch, we here take
the effect of viscous heating in the global evolution of PPDs into account
in addition to the loss of mass and angular momentum by the disc wind. 
We focus in particular on the conditions that create a density structure that
is very different from the structure of classic viscously accreting discs, which may
help solving long-standing problems such as the radial migration of pebbles, 
boulders, and protoplanets.
For this goal, we evaluate the mass-loss rate from the global energetics
of PPDs; the kinetic energy of the vertical outflow is mainly supplied from
the gravitational accretion energy.
This strategy is different from the method adopted by \citet{bai16b}, in which
the mass-loss rate was estimated based on the local profile of magnetically
driven wind with external heating.
A comparison between the two models is provided in Sect. \ref{sec:comp}.

\section{Model}
\label{sec:setup}
\subsection{Basic definitions}
We investigated the time evolution of PPDs with magnetically
driven disc winds. \citet{suz10} solved the evolution of PPDs with
MRI-triggered disc winds under simplified assumptions:  The temperature is
locally constant with time, and the disc wind only contributes to the mass
loss without removing additional angular momentum. 
In this paper, we relaxed these assumptions to treat more realistic evolution
of PPDs.
We considered the heating by viscous accretion \citep{ss73,nn94,hg05}
and the effect of disc wind torque on mass accretion
\citep{bp82,pp92,sal11,bs13b}

Throughout this paper, we assume that each annulus at radial distance $r$
from a central star almost rotates with Keplerian frequency,
$\Omega_{\rm K}$
\begin{equation}
\Omega  \approx \Omega_{\rm K} = \sqrt{\frac{G M_{\star}}{r^{3}}},
\label{eq:Keprot}
\end{equation}
where $G$ is the gravitational constant and $M_{\star}$ is the mass of
the central star. We considered a central star with
solar mass,  $M_{\star}=M_{\odot}$. 
We defined a vertical scale height, $H$, of a disc
\begin{equation}
  H = \frac{\sqrt{2}c_{\rm s}}{\Omega}, 
  \label{eq:sclh}
\end{equation}
where $c_{\rm s}$ is sound speed.  Temperature $T$ and $c_{\rm s}$ are related
through
\begin{equation}
  c_{\rm s}^2 = \frac{k_{\rm B} T} {\mu m_{\rm H}}.
  \label{eq:cs}
\end{equation}
where $k_{\rm B}$ is the Boltzmann constant, $m_{\rm H}$ is the proton mass, and
we assume mean molecular weight, $\mu=2.34$ \citep{hay81}. 
A different definition for the scale height from ours,
$c_{\rm s}/\Omega$ (without $\sqrt{2}$), 
is sometimes used in literatures.

\subsection{Evolution of surface density}
We treated the time evolution of the radial profile of
surface density, $\Sigma=\int dz \rho$, of a disc ($1+1$ D model),
while basic formula transformation is done in cylindrical coordinates,
$(r,\phi,z)$. 
The time evolution of $\Sigma (r)$ can be 
expressed as (see Appendix \ref{sec:apsig} for the derivation)
\begin{displaymath}
\frac{\partial \Sigma}{\partial t} - \frac{1}{r}\frac{\partial}{\partial r}
\left[\frac{2}{r\Omega}\left\{\frac{\partial}{\partial r}\left(r^2 \int dz
  \left(\rho v_r \delta v_{\phi} - \frac{B_rB_{\phi}}{4\pi}\right)\right)
  \right.\right.
\end{displaymath}
\begin{equation}
  \left.\left.
  + r^2 \left(\rho \delta v_{\phi} v_z - \frac{B_{\phi}B_z}{4\pi}\right)_{\rm w}
  \right\}\right]
+ (\rho v_z)_{\rm w} = 0, 
  \label{eq:sgmevlori}
\end{equation}
where $\delta v_{\phi} = v_{\phi} - r\Omega$ is deviation from the background
rotation, and the subscript w stands for disc wind
(see below). 
The $[\cdots]$ parenthesis of the second term represents radial mass flow, 
\begin{displaymath}
-r \Sigma v_r = \frac{2}{r\Omega}\left\{\frac{\partial}{\partial r}
\left(r^2 \int dz\left(\rho v_r \delta v_{\phi}
- \frac{B_rB_{\phi}}{4\pi}\right)\right) \right.
\end{displaymath}
\begin{equation}
\left.
+ r^2 \left(\rho \delta v_{\phi} v_z - \frac{B_{\phi}B_z}{4\pi}\right)_{\rm w}
\right\},
\label{eq:rdmsfx}
\end{equation}
which is derived from the radial balance of angular momentum (Appendix
\ref{sec:apsig}), and the third term of Eq.
  (\ref{eq:sgmevlori}) denotes the mass loss by the disc wind.
The second term consists of the $r\phi$ and $\phi z$
components of Reynolds and Maxwell stresses. 
The ${r\phi}$ component represents the mass accretion (or decretion)
induced by the transport of angular momentum through MHD turbulence.
We used the following parametrization based on
the $\alpha$-prescription introduced by \citet{ss73}:
\begin{equation}
  \int dz \left(\rho v_r \delta v_{\phi} - \frac{B_rB_{\phi}}{4\pi}\right)
  \equiv \int dz \rho \alpha_{r\phi} c_{\rm s}^2 \equiv \Sigma
  \overline{\alpha_{r\phi}} c_{\rm s}^2, 
  \label{eq:rphistress}
\end{equation}
where $\overline{\alpha_{r\phi}}$ is the mass-weighted vertical average of
$\alpha_{r\phi}$.
  $\alpha_{r\phi}$ is a nondimensional parameter normalized by gas
  pressure ($\rho c_{\rm s}^2$) that describes the transport of angular momentum.
  We considered $\alpha_{r\phi}$ to originate from the MHD turbulence induced
  by MRI.
  $\overline{\alpha_{r\phi}}(\lesssim 1)$ depends on physical conditions
  of PPDs, such as the ionization and the strength of poloidal magnetic field; 
  see Sect. \ref{sec:prm} for our adopted values.
  Although we did not separate the contributions from the Reynolds stress
  ($\rho v_r \delta v_{\phi}$) and from the Maxwell stress
  ($-B_r B_{\phi}/4\pi$ $>0$), the latter usually dominates the former
  by a factor of $\sim 4$ in accretion discs with MRI turbulence
  \citep{san04,pes06,haw11} .

$\overline{\alpha_{r\phi}}$ is an effective turbulent viscosity,
  and it is mathematically related to viscosity, $\nu$,
appeared in a hydrodynamical equation, 
\begin{equation}
  \overline{\alpha_{r\phi}} c_{\rm s}^2 = -\nu r \frac{\partial \Omega}{\partial r}
  \approx \frac{3}{2}\nu\Omega, 
  \label{eq:viscosity}
\end{equation}
where the second equality comes from the condition of the Keplerian rotation.
The definition of $\alpha$ is not consistent throughout the literature;
for example, $\nu \approx \alpha_{\rm t} H c_{\rm s}$, is often used conventionally
\citep[e.g.,][]{bh98}.
These two $\alpha$'s are related by
$\alpha_{\rm t} \approx \frac{\sqrt{2}}{3}\overline{\alpha_{r\phi}}$,
where note again that the definition of $H$ (Eq. \ref{eq:sclh}) is
also not consistent in the literatures. 

The $\phi z$ component of the second term of
Eq. (\ref{eq:sgmevlori}) indicates the mass accretion induced by the
angular momentum loss with magnetized disc winds, which was not
taken into account in \citet{suz10}. 
The term of $(\cdots)_{\rm w}$ represents the sum of the angular momentum flux
density carried away by the magnetized outflows from the top and
bottom surfaces of a disc.  While Reynolds
($\rho \delta v_{\phi} v_z$) and Maxwell ($-B_{\phi}B_z/4\pi(>0)$) stresses
contribute to the $\phi z$ stress as well, 
the latter usually dominates in magnetized accretion discs
\citep[e.g.,][]{pp92}, similarly to the $r\phi$ component.
This magnetic braking effect needs to be evaluated in the wind region where it
operates; this is the reason why the subscript w is necessary
in this term. To incorporate the effect of the wind torque
into the 1+1D ($t$--$r$) model, $\alpha_{\phi z}$ needs to be evaluated
by physical quantities at the midplane, 
and we adopted a similar parametrization to the $r\phi$ component, 
\begin{equation}
  \left(\rho \delta v_{\phi} v_z - \frac{B_{\phi} B_z}{4\pi}\right)_{\rm w}
  \equiv (\rho c_{\rm s}^2\alpha_{\phi z})_{\rm w} \equiv
  (\rho c_{\rm s}^2)_{\rm mid} \overline{\alpha_{\phi z}},  
  \label{eq:phizstress}
\end{equation}
where we define nondimensional stress, $\overline{\alpha_{\phi z}}$, normalized
by density, $\rho_{\rm mid} (=\Sigma /(\sqrt{\pi} H))$, at the midplane,


The third term, $(\rho v_z)_{\rm w}$, of Eq. (\ref{eq:sgmevlori}) represents
the sum of the mass loss by the vertical outflows 
from the upper and lower disc surfaces.
\citet{suz10} introduced the nondimensional mass flux normalized
  by the density and the sound speed at the midplane: 
\begin{equation}
  (\rho v_z)_{\rm w} = C_{\rm w}(\rho c_{\rm s})_{\rm mid}. 
  \label{eq:Cw}
\end{equation}
We model $C_{\rm w}$ in Sect. \ref{sec:Cw}.

Substituting Eqs. (\ref{eq:rphistress}), (\ref{eq:phizstress}), and
(\ref{eq:Cw}) into Eq. (\ref{eq:sgmevlori}), we finally have 
\begin{displaymath}
\frac{\partial \Sigma}{\partial t} - \frac{1}{r}\frac{\partial}{\partial r}
\left[\frac{2}{r\Omega}\left\{\frac{\partial}{\partial r}(r^2 \Sigma
  \overline{\alpha_{r\phi}}c_{\rm s}^2) + r^2 \overline{\alpha_{\phi z}}
  (\rho c_{\rm s}^2)_{\rm mid} \right\}\right]
\end{displaymath}
\begin{equation}
+ C_{\rm w}(\rho c_{\rm s})_{\rm mid} = 0. 
  \label{eq:sgmevl}
\end{equation}
We solved this equation for different sets of the three parameters,
$\overline{\alpha_{r\phi}}$, $\overline{\alpha_{\phi z}}$, and $C_{\rm w}$. 
We note that \citet{bai16b} recently derived essentially the same equation in a
different form using mass-loss rate and mass accretion rate instead of
the above three-dimensionless parameters. 

\subsection{Mass-loss rate by disc winds: Energetics.}
\label{sec:Cw}
We assumed that the energy of the disc wind originates from
gravitational accretion. 
Then, the mass flux of the disc wind, $C_{\rm w}$, is constrained by
$\overline{\alpha_{r\phi}}$ and $\overline{\alpha_{\phi z}}$.
A starting point for this energetics constraint is the conservation equation
of total MHD energy \citep[e.g.,][]{bh98},
\begin{displaymath}
\frac{\partial}{\partial t}\left[\frac{1}{2}\rho v^2 + \rho \Phi
  +\frac{p}{\gamma -1} + \frac{B^2}{8\pi} \right]
+\mbf{\nabla\cdot}\left[\mbf{v}\left(\frac{1}{2}\rho v^2 + \rho \Phi
    +\frac{\gamma p}{\gamma -1} \right) \right.
\end{displaymath}
\begin{equation}
  \left. +\frac{\mbf{B}}{4\pi}\times
    (\mbf{v}\times\mbf{B}) + \mbf{F}_{\rm ot}\right] = 0, 
\label{eq:totengMHD}
\end{equation}
where $p$ is gas pressure, $\gamma$ is a ratio of specific heats,
$\Phi=-GM_{\star}/r = -r^2\Omega_{\rm K}^2\approx -r^2\Omega^2$ is the
gravitational potential
by a central star, and $\mbf{F}_{\rm ot}$ is other contributions to energy flux
in addition to the MHD energy, such as thermal conduction and radiative
heating or cooling.
We considered thin discs with nearly Keplerian rotation
(Eq. \ref{eq:Keprot}), and hence, we can assume $r\Omega \gg
v_r, \delta v_{\phi}, v_z, c_s, B/\sqrt{4\pi\rho}$, and safely neglect the
terms concerning gas pressure.
Leaving dominant terms in Eq. (\ref{eq:totengMHD}) we finally obtained
an approximated energy equation
as (Appendix \ref{sec:dereng}; Eq. \ref{eq:totezint2})
\begin{displaymath}
\frac{\partial}{\partial t}\left(-\Sigma \frac{r^2\Omega^2}{2}\right) 
+\frac{1}{r}\frac{\partial}{\partial r}\left[r\Omega
  \left\{\frac{\partial}{\partial r}(r^2 \Sigma
\overline{\alpha_{r\phi}}c_{\rm s}^2) + r^2 \overline{\alpha_{\phi z}}
(\rho c_{\rm s}^2)_{\rm mid} \right\} \right.
\end{displaymath}
\begin{equation}
  \left.
  + r^2\Omega\Sigma\overline{\alpha_{r \phi}}
  c_{\rm s}^2\right] + (\rho v_z)_{\rm w} E_{\rm w}+ F_{\rm rad} =0, 
\label{eq:totengring}
\end{equation}
where $E_{\rm w}$ is the specific total energy of the gas in the disc wind;
$(\rho v_z)_{\rm w}E_{\rm w}$ is the energy carried away by the disc wind. 
$F_{\rm rad}$ is radiation loss from the top and bottom surfaces, 
\begin{equation}
  F_{\rm rad} = 2\sigma_{\rm SB}T_{\rm surf}^4,
  \label{eq:SB}
\end{equation}
where $\sigma_{\rm SB}$ is the Stefan-Boltzmann constant and $T_{\rm surf}$ is
the temperature at the disc surfaces.
We here neglected the energy gain by the irradiation from a central star
\citep{kus70,dul02,dav05} and other external sources.
The effect of stellar irradiation was taken into account later when we
estimated the temperature.

Equation (\ref{eq:totengring}) contains two terms with
$\overline{\alpha_{r\phi}}$; the first term in $\{\cdots\}$ denotes the liberated
gravitational energy by mass accretion, and second term outside $\{\cdots\}$
represents heating by turbulent dissipation, which phenomenologically
corresponds to viscous heating.  
The wind torque, $\overline{\alpha_{\phi z}}$, does not contribute to this
effective viscous heating because the disc wind
does not transport angular momentum within the disc but simply removes it,
although $\overline{\alpha_{\phi z}}$ contributes to the mass accretion. 

Using Eq. (\ref{eq:sgmevl}), we can eliminate the time derivative term
of Eq. (\ref{eq:totengring}) to derive an energetics constraint on the disc
wind (Appendix \ref{sec:dereng}):
\begin{displaymath}
  (\rho v_z)_{\rm w}\left(E_{\rm w}+\frac{r^2\Omega^2}{2}\right) + F_{\rm rad}
\end{displaymath}
\begin{eqnarray}
  &=& \frac{\Omega}{r}\left[\frac{\partial}{\partial r}(r^2 \Sigma
    \overline{\alpha_{r\phi}}c_{\rm s}^2) + r^2
    \overline{\alpha_{\phi z}}(\rho c_{\rm s}^2)_{\rm mid}\right] \nonumber \\
  & &\hspace{1cm}-\frac{1}{r}\frac{\partial}{\partial r}(r^2 \Sigma\Omega
  \overline{\alpha_{r\phi}}c_{\rm s}^2)
  \label{eq:engcnt1}
  \\
  &=&\frac{3}{2}\Omega\Sigma\overline{\alpha_{r\phi}}c_{\rm s}^2 + r\Omega
  \overline{\alpha_{\phi z}}(\rho c_{\rm s}^2)_{\rm mid} 
  \label{eq:engcnt2}
\end{eqnarray}
The physical meaning of Eq. (\ref{eq:engcnt1}) is that
the energy carried away by disc winds (first term on the left-hand side;
l.h.s. hereafter) and radiation (second term on the l.h.s.) is determined by
the gravitational energy liberated by accretion (first term on the right-hand
side; r.h.s. hereafter) and effective
viscous heating (second term on the r.h.s.).  We used
the Keplerian rotation (Equation \ref{eq:Keprot}) to transform 
Eq. (\ref{eq:engcnt1}) to Eq. (\ref{eq:engcnt2}).
The term with $\overline{\alpha_{r\phi}}$ includes contributions from
the gravitational accretion and from the effective viscous heating. 

\citet{suz10}
  assumed that $E_{\rm w}\ge \frac{3}{2} v_z^2$ is the condition to drive
  the vertical outflow to a large distance \citep[Eq. 22 of ][]{suz10}. 
  However,  we adopt $E_{\rm w}\ge 0$, because this is the sufficient
  condition for the wind material to reach $z\Rightarrow \infty$
  ($v_z^2>0$ in Eq. \ref{eq:Edwinfty}).
Following this consideration, we derived the mass flux
of the disc wind that satisfies the energetics constraint with $E_{\rm w}=0$
from Eqs. (\ref{eq:engcnt1}) and (\ref{eq:engcnt2})  
in a nondimensional form: 
\begin{displaymath}
  C_{\rm w,e} +  \frac{2 F_{\rm rad}}{r^2\Omega^2 (\rho c_{\rm s})_{\rm mid}}
\end{displaymath}
\vspace{-0.5cm}
\begin{eqnarray}
  &=& \frac{2}{r^3\Omega(\rho c_{\rm s})_{\rm mid}}\frac{\partial}{\partial r}
  (r^2 \Sigma\overline{\alpha_{r\phi}}c_{\rm s}^2)
  + \frac{2c_{\rm s}}{r\Omega}\overline{\alpha_{\phi z}} \nonumber \\
  & &- \frac{2}{r^3\Omega^2(\rho c_{\rm s})_{\rm mid}}\frac{\partial}{\partial r}
  (r^2 \Sigma\Omega\overline{\alpha_{r\phi}}c_{\rm s}^2)
  \label{eq:Cwcn1}
  \\
  &=&\frac{3\sqrt{2\pi}c_{\rm s}^2}{r^2\Omega^2}\overline{\alpha_{r\phi}}
  + \frac{2c_{\rm s}}{r\Omega}\overline{\alpha_{\phi z}},
  \label{eq:Cwcn2}
  \\
  &=& 3\sqrt{\pi/2}h^2 \overline{\alpha_{r\phi}}
  + \sqrt{2}h\overline{\alpha_{\phi z}} \nonumber
\end{eqnarray}
where $C_{\rm w,e}$ stands for the mass flux constrained by the energetics. 
We here used $\Sigma \Omega = \sqrt{2\pi}(\rho c_{\rm s})_{\rm mid}$,
and for the last equality we introduced an aspect ratio,
$h\equiv H/r=\sqrt{2}c_{\rm s}/r\Omega$.

It is crucial 
to determine the fractions of the energy transferred to the disc
winds (first term on the l.h.s. of Eq. \ref{eq:Cwcn1}) and to the
radiation loss (second term).
Following the standard accretion disc model \citep{ss73}, the
available energy from the viscous accretion is transferred to the radiation.
In the magnetocentrifugal driven wind model \citep{bp82}, the angular
momentum carried by disc winds is directly related to the wind mass-loss
rate. Based on these models, we may infer that the
$\overline{\alpha_{r\phi}}$ term in Eq. (\ref{eq:Cwcn2}) regulates
the $F_{\rm rad}$ term and the $\overline{\alpha_{\phi z}}$ term determines
$C_{\rm w,e}$. However, the situation is not this simple, because disc 
winds can be launched solely by the $\overline{\alpha_{r\phi}}$ term, which
was shown by local shearing box simulations with zero-wind torque,
$\overline{\alpha_{\phi z}}=0$ \citep{si09}. MRI excites MHD turbulence and
the associated Poynting flux drives vertical outflows. The original energy
source in this mechanism is the energy released by the gravitational
accretion.

Despite these complicated problems, 
we adopted two different strategies to determine $C_{\rm w,e}$
and $F_{\rm rad}$ in this paper.
The first strategy is that $F_{\rm rad}$ is equal to the effective viscous
heating and all the liberated gravitational energy is transferred to the
disc winds. The first corresponds to the first line on the right-hand side of
Eq. (\ref{eq:Cwcn1}), and the second corresponds to the second line, and
then,
\begin{eqnarray}
  C_{\rm w,e}\hspace{-0.3cm}&=&\hspace{-0.3cm}\max\left(\frac{2}{r^3\Omega(\rho c_{\rm s})_{\rm mid}}
  \frac{\partial}{\partial r}(r^2 \Sigma\overline{\alpha_{r\phi}}c_{\rm s}^2)
  + \frac{2c_{\rm s}}{r\Omega}\overline{\alpha_{\phi z}},0\right)
  \label{eq:DWlossCw}
  \\
  F_{\rm rad}\hspace{-0.3cm}&=&\hspace{-0.3cm}\max\left(-\frac{1}{r}\frac{\partial}{\partial r}
  (r^2 \Sigma\Omega\overline{\alpha_{r\phi}}c_{\rm s}^2),0\right), 
  \label{eq:DWlossFrad}
\end{eqnarray}
where we avoided negative values of $C_{\rm w,e}$ and $F_{\rm rad}$.

In the second choice we left the uncertainty to a parameter,
$\epsilon_{\rm rad}$, that determines the fractional energy to the
radiation loss: 
\begin{eqnarray}
  C_{\rm w,e} &=& (1-\epsilon_{\rm rad})\left[\frac{
      3\sqrt{2\pi}c_{\rm s}^2}{r^2\Omega^2}\overline{\alpha_{r\phi}}
    + \frac{2c_{\rm s}}{r\Omega}\overline{\alpha_{\phi z}}\right]
  \label{eq:RadlossCw}
  \\
  &=& (1-\epsilon_{\rm rad})\left[3\sqrt{\pi/2}h^2 \overline{\alpha_{r\phi}}
  + \sqrt{2}h\overline{\alpha_{\phi z}}\right] \nonumber
\end{eqnarray}
\begin{equation}
  F_{\rm rad} = \epsilon_{\rm rad} \left[\frac{3}{2}\Omega\Sigma
    \overline{\alpha_{r\phi}}c_{\rm s}^2 + r\Omega
  \overline{\alpha_{\phi z}}(\rho c_{\rm s}^2)_{\rm mid}\right].
  \label{eq:RadlossFrad}
\end{equation}
Since the first method is an extreme limit for the maximum disc wind flux,
we sought the other extreme limit of great radiation loss in the
second method; we adopted $\epsilon_{\rm rad}=0.9$.
We name the first case (Eqs. \ref{eq:DWlossCw} \& \ref{eq:DWlossFrad})
strong DW and the second case (Eqs. \ref{eq:RadlossCw} \&
\ref{eq:RadlossFrad} with $\epsilon_{\rm rad}=0.9$) weak DW
from here on; DW stands for disc wind.

On the other hand,
local MHD shearing box simulations also give the mass flux of disc winds  
\citep{si09,suz10}.
We constrained the mass flux of the local simulations, $C_{\rm w,0}$,
by the energetics of the global accretion to give the $C_{\rm w}$ that we 
use in our calculations, 
\begin{equation}
  C_{\rm w} = \min(C_{\rm w,0},C_{\rm w,e}), 
  \label{eq:Cwe,0}
\end{equation}
where the adopted $C_{\rm w,0}$ is presented in Sect. \ref{sec:prm}.


\subsection{Temperature: viscous heating \& radiative equilibrium.}
By referring to the terms concerning $\overline{\alpha_{r\phi}}$ in Eq.
(\ref{eq:engcnt1}), the viscous heating rate can be scaled as $\sim \Sigma \Omega
c_{\rm s}^2$. Since $\Sigma$ decreases with $t$ and $\Omega c_{\rm s}^2$
has a negative dependence on $r$, 
the viscous heating is anticipated to play a primary role in determining
the temperature in the inner region ($\lesssim 10$ au) and
at the early stage of the evolution of a PPD.
As $\Sigma$ decreases with the dispersal of the gas component, the disc evolves
passively by the illumination from the central star. 
A number of works have been 
published that treat this problem with detailed models that include
viscous heating and stellar irradiation
\citep[e.g.,][]{gl07,oka11,bit15}.

If 
the viscous heating is more effective 
in a PPD than stellar irradiation, then the temperature at the midplane,
$T_{\rm mid}$, will be higher than $T_{\rm surf}$ in Eq. (\ref{eq:SB}).
On the other hand, if the viscous heating is ineffective and the stellar
irradiation dominates, then $T_{\rm surf}$ will be higher than
$T_{\rm mid}$.
The radiative transfer needs to be solved to determine the vertical
temperature profile.  
However, since our main focus here 
is to investigate the roles of magnetically driven disc winds, we adopt
the simple prescription for the temperature that was
introduced by \citet{nn94}. We defined $T_{\rm vis}$ as the
temperature at the midplane determined by viscous heating, 
\begin{equation}
  2\sigma_{\rm SB} T_{\rm vis}^4 = \left(\frac{3}{8}\tau_{\rm R}
  +\frac{1}{2\tau_{\rm P}}\right) F_{\rm rad}
  \label{eq:visT}
\end{equation}
where $\tau_{\rm R}$ and $\tau_{\rm P}$ are the Rosseland mean optical depth
and the Planck mean optical depth measured at the midplane.
$\tau_{\rm R}$ is estimated from the surface density and the Rosseland mean
opacity, $\kappa_{\rm R}$,
\citep{hg05} as
\begin{equation}
  \tau_{\rm R} = \kappa_{\rm R}\Sigma  / 2, 
\end{equation}
where we use 
\begin{eqnarray}
  \kappa_{\rm R} = \left\{
  \begin{array}{ll}
    4.5\left(\frac{T}{150{\rm K}}\right)^2 {\rm cm^{2}g^{-1}} & : T<150\; {\rm K}\\
    4.5\;{\rm cm^{2}g^{-1}} & : 150\;{\rm K}\le T\le 1500\; {\rm K}\\
    0\;{\rm cm^{2}g^{-1}} & : T>1500\; {\rm K}
  \end{array}
  \right. ,
  \label{eq:kappaR}
\end{eqnarray}
based on the opacity of dust grains \citep[][see also \citealt{bail15}]{nn94}.
The Planck mean optical depth can be approximated as
\begin{equation}
  \tau_{\rm P}=\max(2.4\tau_{\rm R},0.5)
\end{equation}
\citep{nn94,hg05}, where we give the lower bound on $\tau_{\rm P}$
to obtain the pre-factor of Eq. (\ref{eq:visT}),
$\left(\frac{3}{8}\tau_{\rm R}+\frac{1}{2\tau_{\rm P}}\right)\Rightarrow 1$,
for the optically thin limit. 

We can also define the temperature under the radiation
equilibrium, which is determined by the irradiation from the central star, 
\begin{equation}
  T_{\rm req} = T_{\rm 1au} \left(\frac{r}{1{\rm au}}\right)^{p}. 
  \label{eq:treq}
\end{equation}
We adopted $T_{\rm 1au}=280$ K and $p=-1/2$ based on the simple radiative
equilibrium for the original minimum mass solar nebula (MMSN; hereafter)
\citep{hay81,hay85}. We note that a slightly different scaling is derived,
when the geometry of a flared disc is taken into account \citep{cg97,cy10}.

When a PPD becomes optically thin and the viscous heating is
ineffective, not only $T_{\rm surf}$ but also $T_{\rm mid}$ approaches 
$T_{\rm req}$. To take both viscous heating and stellar irradiation into account,
we tool the sum of these two temperatures,
\begin{equation}
  T^4 = T_{\rm vis}^4 + T_{\rm req}^4,
  \label{eq:Tselect}
\end{equation}
for the representative $z$-averaged temperature, $T$, to estimate $c_{\rm s}$
in Equation (\ref{eq:cs}).


\subsection{Initial and boundary conditions}
We calculated the evolution of $\Sigma$ of the initial profile, 
$\propto r^{-3/2}$ \citep{hay81,hay85}, with a cut-off radius $r_{\rm cut}$
\begin{equation}
  \Sigma_{\rm int} = \Sigma_{1{\rm au}}
  \left(\frac{r}{\rm 1\; au}\right)^{-3/2}
  \exp\left(-\frac{r}{r_{\rm cut}}\right). 
  \label{eq:Sgminit}
\end{equation}
The original MMSN by \citet[][]{hay81} considered
$\Sigma_{1{\rm au}}=1.7\times 10^3$g cm$^{-2}$ with a sharp cut-off at $36$ au, 
which gives the initial disc mass, $M_{\rm disc,int} = 0.013M_{\odot}$.
We adopted a ten times larger $\Sigma_{1{\rm au}}=1.7\times 10^4$g cm$^{-2}$ but
slightly smaller $r_{\rm cut}=30$ au in this paper, which gives
$M_{\rm disc,int}=0.11 M_{\odot}$.
Mass accretion rates are observationally obtained as a function of time
\citep{gul98,har98,ric10,man16}, which corresponds to the age of the central
stars, while the MMSN corresponds to a late stage of the evolution.
Therefore, we chose the massive initial disc to directly compare our
results to these observations. 

We solved Eq. (\ref{eq:sgmevl}) to track the time evolution of $\Sigma$
in the region from $r_{\rm in}=0.01$ au to $r_{\rm out}=10^4$ au with grid spacing,
$\Delta r \propto \sqrt{r}$.
At the inner and outer boundaries, $r=r_{\rm in}$ and $=r_{\rm out}$, we imposed
$\frac{\partial}{\partial r}(\Sigma r^{3/2})=0$, which corresponds to the
zero-torque boundary condition \citep{lp74}; the $\overline{\alpha_{r\phi}}$
term in Eq. (\ref{eq:sgmevl}), $\frac{\partial}{\partial r}(r^2 \Sigma
\overline{\alpha_{r\phi}}c_{\rm s}^2)$, is zero for a constant
$\overline{\alpha_{r\phi}}$ and 
$c_{\rm s}^2\propto r^{-1/2}$ (Eq. \ref{eq:treq} with $p=-1/2$). 

\subsection{Parameters}
\label{sec:prm}
The free parameters of our model are turbulent viscosity,
$\overline{\alpha_{r\phi}}$, disc wind mass flux, $C_{\rm w,0}$, and
disc wind torque, $\overline{\alpha_{\phi z}}$. 
We would like to note that, although we here call
$\overline{\alpha_{r\phi}}$ turbulent viscosity, large-scale magnetic
fields possibly contribute to $\overline{\alpha_{r\phi}}$ in realistic situations
\citep{ts08,joh09}.

\subsubsection{Turbulent viscosity -- $\overline{\alpha_{r\phi}}$}
We compared two cases with spatially uniform
$\overline{\alpha_{r\phi}}=8\times 10^{-3}$, and
$8\times 10^{-5}$. $\overline{\alpha_{r\phi}}=8\times 10^{-3}$ 
was adopted from the result of local shearing box MHD simulations with
sufficient ionization \citep[][see also e.g. \citealt{san04,sai13}]{suz10}
in which MHD turbulence is fully developed by the MRI.
When the ionization is not sufficient and non-ideal
MHD effects such as resistivity, Hall diffusion, and ambipolar diffusion are
important, a magnetically inactive dead zone forms \citep{gam96} and
$\overline{\alpha_{r\phi}}$ is smaller
\citep{san98,ll07,sim11,oh11,flo12,gre15}. We adopted $\overline{\alpha_{r\phi}}
=8\times 10^{-5}$ for such MRI-inactive circumstances. 
Although we assumed constant
$\overline{\alpha_{r\phi}}$ for simplicity, $\overline{\alpha_{r\phi}}$ would
be spatially dependent on $r$ and evolve with time in realistic situations,
because a dead zone generally forms only in the inner region and
its size shrinks with time \citep[e.g.,][]{san00,suz10,dzy13}.
For future elaborate studies, we need to take this spatially and
time-dependent $\overline{\alpha_{r\phi}}$ into account.

\subsubsection{Disc wind mass flux -- $C_{\rm w,0}$}
The mass flux of disc winds, $C_{\rm w,0}$, was also adopted from the local
simulations.
$C_{\rm w,0}$ is controlled by the density at the wind onset region,
which is located at the upper regions where the magnetic energy becomes
comparable to the thermal energy. For the MRI turbulence, depending on the net
vertical magnetic field, the density at the wind footpoint is
$\approx 10^{-5}-10^{-4}$ times the density at the midplane, which
gives $C_{\rm w,0}\approx 10^{-5} - 10^{-4}$.
Here, we add a note of caution: the local simulations might overestimate
the mass-loss rate of the disc winds because the returning mass to the
simulation box cannot be properly taken into account.
\citet{suz10} reported that the mass flux is reduced by a factor of
2-3 in simulations with a larger vertical box size. \citet{fro13} also
pointed
out that the reduction factor could be as large as $\sim 10$, but their
numerical scheme and other detailed set-up were different from those used in
\citet{suz10}.
These results show that we must choose $C_{\rm w,0}$ carefully from the local
simulations.

When we take the face value of the local simulations assuming the ideal
MHD condition, $C_{\rm w,0}\approx 4\times 10^{-5}$ for the weak vertical
magnetic field \citep{si09}. 
We here set a more conservative value, $C_{\rm w,0}=2\times 10^{-5}$,
for the MRI-active cases
with $\overline{\alpha_{r\phi}}=8\times 10^{-3}$. If a dead
zone is formed, then the mass flux of the disc winds is slightly
reduced, but it does not become as low as $\overline{\alpha_{r\phi}}$ 
because the disc winds are driven
from the surface regions with sufficient ionization; $C_{\rm w,0}$ is only
moderately weakened by a factor of a few.
We adopted $C_{\rm w,0}=1\times 10^{-5}$ for $\overline{\alpha_{r\phi}}
=8\times 10^{-5}$. 
Moreover, the actual mass flux,
$C_{\rm w}$, is constrained by the energetics, Eq. (\ref{eq:Cwe,0}).
We also assumes, in the same way as $\overline{\alpha_{r\phi}}$,
constant $C_{\rm w,0}$ for simplicity. While in realistic situations
it would depend on $r$ and vary with time, it does not change as much as
$\overline{\alpha_{r\phi}}$. 

\subsubsection{Disc wind torque -- $\overline{\alpha_{\phi z}}$}
We tested two types of the parametrization for the wind torque: 
(i) constant $\overline{\alpha_{\phi z}}=1\times 10^{-4}$, and (ii) density
dependent with a cap,
\begin{equation}
  \overline{\alpha_{\phi z}}=\min\left(10^{-5}\left(\frac{\Sigma}{\Sigma_{\rm int}}
  \right)^{-0.66},1\right).
  \label{eq:apzddp}
\end{equation}
  We name (i) constant torque and (ii) $\Sigma$-dependent torque from now on.
$\overline{\alpha_{\phi z}}$ was estimated by local MHD simulations by
\citet{bai13},
who reported $\overline{\alpha_{\phi z}}\sim 10^{-5} - 10^{-3}$ with a positive
dependence on the strength of the net vertical magnetic field,
$\overline{\alpha_{\phi z}}\propto (B_z^2 /8\pi
(\rho c_{\rm s}^2)_{\rm mid})^{0.66}$. $\rho_{\rm mid}$ is proportional to
$\Sigma$, while $B_z$ is determined by the inward dragging and outward
diffusion of magnetic flux
\citep[][see also Subsection \ref{sec:comp}]{lub94,oku14,go14}.
If $B_z$ decreases with the dispersal of gas (decrease of $\Sigma$), then
$\overline{\alpha_{\phi z}}$ will stay approximately constant, which
corresponds to
(i) constant torque; if $B_z$ does not decrease that much,
then $\overline{\alpha_{\phi z}}$ has a negative dependence on $\Sigma$ and will
increase with time, which corresponds to (ii) $\Sigma$-dependent torque.
We tested these two extreme limits for the effect of the wind
torque affected by the evolution of the vertical magnetic flux.

\section{Results}
\label{sec:res}
In this section, we present the properties of the time evolution of PPDs in the
MRI-active and MRI-inactive conditions. 

\subsection{MRI-active cases}

\begin{table*} 
  \begin{center}
  \caption{Parameters for MRI-active cases
    \label{tab:MRIact}
    }
  \begin{tabular}{|c||c|c|c|c|}
    \hline
    Case & $\overline{\alpha_{r\phi}}$ & $C_{\rm w,0}$ & $\overline{\alpha_{\phi z}}$ & Energetics\\
    \hline
    \hline
    Strong DW + $\Sigma$-dependent torque & $8\times 10^{-3}$ & $2\times 10^{-5}$ &
    $10^{-5}(\Sigma/\Sigma_{\rm int})^{-0.66}$
    & Eqs.(\ref{eq:DWlossCw}) and (\ref{eq:DWlossFrad})\\
    Strong DW + zero-torque & $8\times 10^{-3}$ & $2\times 10^{-5}$ & 0
    & Eqs.(\ref{eq:DWlossCw}) and (\ref{eq:DWlossFrad})\\
    Weak DW + zero-torque & $8\times 10^{-3}$ & $2\times 10^{-5}$ & 0
    & Eqs.(\ref{eq:RadlossCw}) and (\ref{eq:RadlossFrad}) with
    $\epsilon_{\rm rad}=0.9$\\
    No DW & $8\times 10^{-3}$ & 0 & 0
    & Eqs.(\ref{eq:RadlossCw}) and (\ref{eq:RadlossFrad}) with
    $\epsilon_{\rm rad}=1$\\
    \hline
  \end{tabular}
  \end{center}
\end{table*}
    
\begin{figure}
  \begin{center}
    \includegraphics[width=0.46\textwidth]{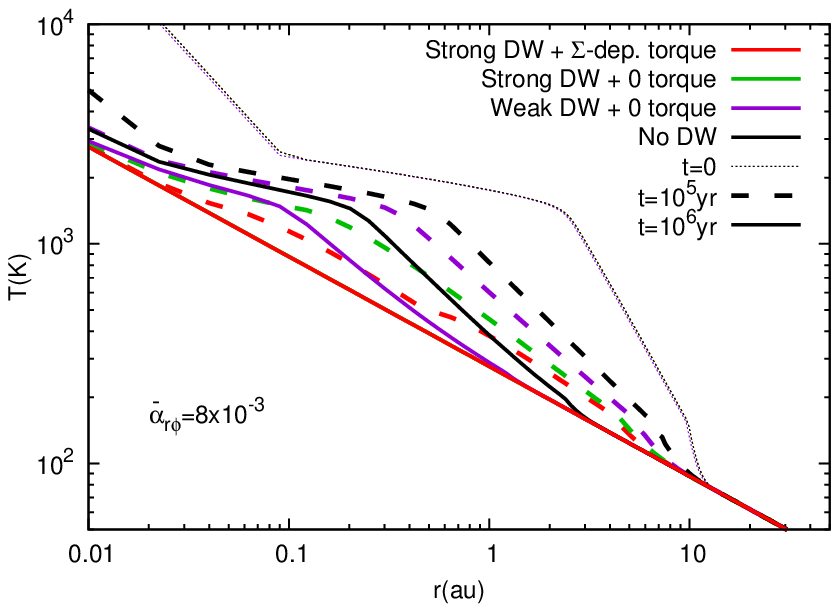}\\
    \includegraphics[width=0.47\textwidth]{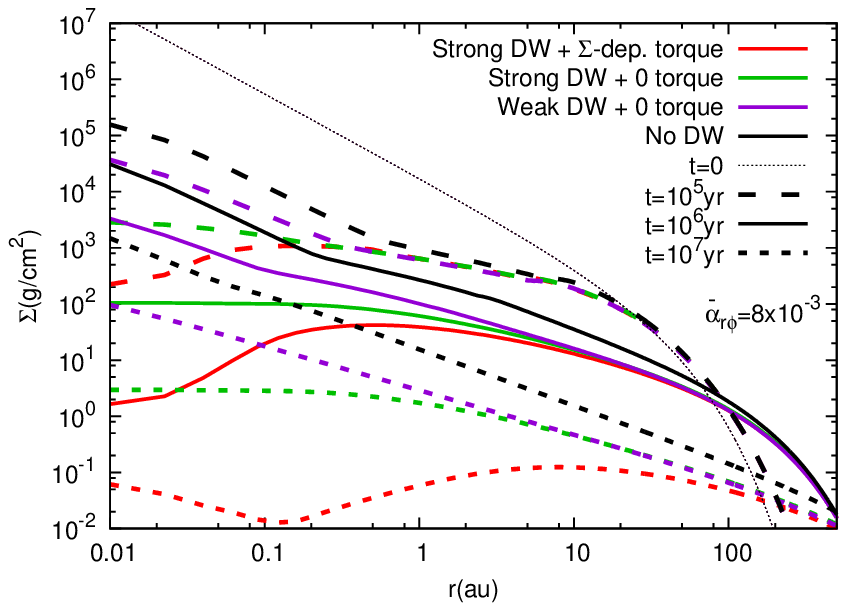}
  \end{center}
  \caption{Comparison of time evolutions of four MRI-active PPDs with 
    $\overline{\alpha_{r\phi}}=8\times 10^{-3}$. The four cases are
    (i) strong DW + $\Sigma$-dependent torque (red), (ii) strong DW + zero-torque
    (green), (iii) weak DW + zero-torque (purple), and (iv) no DW (black), 
    summarized in Table \ref{tab:MRIact}. 
    {\it Top}: Radial profiles 
    of temperatures, $T$, at $t=0$ (dotted lines), $10^5$ (dashed lines),
    and $10^6$ (solid lines) years. We note that the initial temperatures
    of the four cases are almost the same and that the red and green solid
    lines at $t=10^6$ years overlap at $T=T_{\rm req}$
    (Eq. \ref{eq:treq}).
    {\it Bottom}: Radial profiles of surface densities, $\Sigma$, at $t=0$
    (dotted lines), $10^5$ (long dashed lines), $10^6$ (solid lines), and
    $10^7$ (short dashed lines) years.
    We note that the radial range of the top panel is more zoomed-in
    than the radial range of the bottom panel. 
    \label{fig:profile83}
  }
\end{figure}

In this subsection we show results of four cases of MRI-active
  PPDs, which are summarized in Table \ref{tab:MRIact}.
  The first three cases take disc winds into account. The magnetic braking
  by the disc winds is only considered in the first case. The last case
  (no DW) does not take disc winds into account by substituting
  $\epsilon_{\rm rad}=1$ and $C_{\rm w,0}=0$ in Eqs. (\ref{eq:RadlossCw})
  -- (\ref{eq:Cwe,0}).

Figure \ref{fig:profile83} compares radial profiles of $T$ and $\Sigma$
of these four cases.
The top panel compares the evolution of the temperatures of these four cases.
The initial temperature profiles in 0.1 au $\lesssim r \lesssim$ 5 au,
are kept more or less constant $\lesssim 1500-2500$ K
because dust grains sublimate and the opacity drops above that temperature
\citep[Equation \ref{eq:kappaR}; see also][]{bail15}.
Furthermore, the initial profiles are almost the same for 
the four cases,  
except for different energetics constraints on
$C_{\rm w}$ and wind torques, $\overline{\alpha_{\phi z}}$. In particular,
the weak DW case (adopting Eqs. \ref{eq:RadlossCw} and
\ref{eq:RadlossFrad}; purple dotted line)
gives a very similar profile to those of the strong DW
cases (adopting Eqs. \ref{eq:DWlossCw} and \ref{eq:DWlossFrad}; red and
green dotted line), which needs explanation. 
In the inner region, $\lesssim 10$ au, $T\approx T_{\rm vis}$
(Eq. \ref{eq:Tselect}) in these cases, and then $T$ is mainly determined
from $F_{\rm rad}$ by Eq. (\ref{eq:visT}).
Recalling $\Sigma_{\rm int}\propto r^{-3/2}$, we derive
$-\frac{1}{r}\frac{\partial}{\partial r}(r^2 \Sigma \Omega
\overline{\alpha_{r\phi}} c_{\rm s}^2)
\approx \frac{3}{2}\Sigma \Omega \overline{\alpha_{r\phi}} c_{\rm s}^2$
for $c_{\rm s}^2 \sim r^{-1/2}$.
Since the $\overline{\alpha_{\phi z}}$($0$ or $=10^{-5})$ term in Eq.
(\ref{eq:RadlossFrad}) is negligible in comparison to
the $\overline{\alpha_{r\phi}}
(=8\times 10^{-3})$ term, both strong DW and weak DW conditions
give similar $F_{\rm rad}$ in Eqs. (\ref{eq:DWlossFrad}) and
(\ref{eq:RadlossFrad}), and accordingly, the initial temperatures of
these cases are similar each other.

In the no DW case (black lines) the viscous heating region
($T_{\rm vis}>T_{\rm req}$) survives until a later time although its size shrinks.
In contrast, the temperatures decrease more rapidly in the other cases with
disc winds. 
In the two strong DW cases (red and green lines), 
the temperatures are mainly determined by $T_{\rm req}$ in the entire region after
$t\gtrsim 10^6$ years because the surface densities decrease rapidly by the
disc winds in the inner region to give $T_{\rm req} \gg T_{\rm vis}$, while
$T_{\rm vis}$ is no longer negligible in the weak DW case (purple lines) at
$t=10^6$ years.

\begin{figure} 
  \begin{center}
    \includegraphics[width=0.45\textwidth]{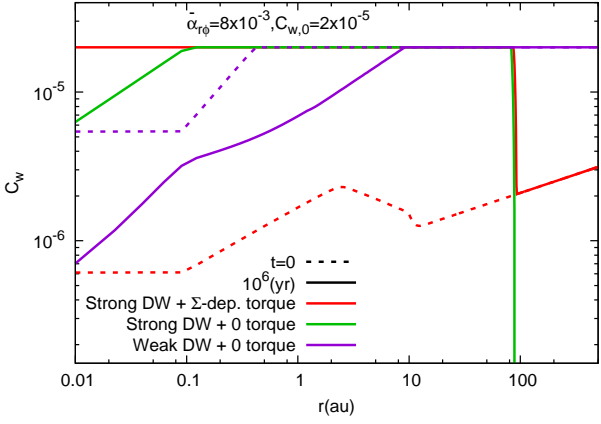}
  \end{center}
  \caption{Comparison of nondimensional mass flux of disc winds, $C_{\rm w}$,
    of the three MRI-active cases except for the no DW case in
    Table \ref{tab:MRIact}.
    at $t=0$ (dotted lines) and $10^6$ years (solid lines).
    \label{fig:Cw83}
  }
\end{figure}

The bottom panel of Fig. \ref{fig:profile83} compares the evolution of
the surface densities.
The disc winds reduce $\Sigma$ particularly in small $r$
regions \citep{suz10}.
A comparison between the two zero-torque cases 
(green and purple lines) shows the difference between the strong DW
and weak DW conditions.
As expected, the strong DW case shows faster decrease of $\Sigma$ 
because of the higher disc wind mass flux, $C_{\rm w}$, which is shown in Fig.
\ref{fig:Cw83}.  
At $t=0$ the strong DW case 
(green dotted line) gives quite small $C_{\rm w}\approx 0$ below the displayed
range of Fig. \ref{fig:Cw83} because $\frac{\partial}{\partial r}
(r^2\Sigma \overline{\alpha_{r\phi}}c_{\rm s}^2)\approx 0$ for $\Sigma_{\rm int}
\propto r^{-3/2}$ in Eq. (\ref{eq:DWlossCw}).
However, as $\Sigma$ decreases in an inside-out manner 
  and the $\Sigma$ profile changes, $C_{\rm w}$ increases
and at $t=10^6$ years this case (green solid line) yields larger $C_{\rm w}$ than the weak DW
case (purple solid line), in which $C_{\rm w}$ instead decreases with time owing
to the decrease in temperature ($\propto c_{\rm s}^2$; Equation
\ref{eq:RadlossFrad}). We note that $C_{\rm w}=0$ in the outer region, $r>90$ au,
of the strong DW + zero-torque case because the gas moves outward
($\frac{\partial}{\partial r}(r^2\Sigma \overline{\alpha_{r\phi}}c_{\rm s}^2)< 0$
in Eq. (\ref{eq:DWlossCw})) in the outer region and
the gravitation energy is not released. In realistic situations, however,
a moderate level of external heating by stellar irradiation or other sources
would cause disc winds to be launched by relaxing the energetics constraint
(see Sect. \ref{sec:unc}),
because the gas is only weakly bound by the gravity in the outer region.

The non-zero wind torque 
also reduces $\Sigma$ faster
(red lines in the bottom panel of Fig. \ref{fig:profile83}) by
the enhanced accretion and disc wind mass-loss. A comparison between
the red and green lines in Fig. \ref{fig:Cw83} indicates
that the removal of angular momentum by the $\phi z$ stress 
additionally contributes to the gravitational
energy by the accretion to enhance $C_{\rm w}$ (Eq. \ref{eq:DWlossCw}). 
As a result, $C_{\rm w}$ is not constrained by the energetics, $C_{\rm w,e}$,
in the almost entire region but is determined by
$C_{\rm w,0}(=2\times 10^{-5})$ at $t=10^6$ years (red solid lines). 
The constant $C_{\rm w}=C_{\rm w,0}$ implies faster dispersal of $\Sigma$
for smaller $r$ because the mass-loss timescale becomes proportional to the
Keplerian time \citep{suz10,ogi15a,ogi15b}, and the slope of $\Sigma$
is positive in the inner region.
The slope of $\Sigma$ is again negative in the very
inner region, $r<0.1$ au, at later time, $t\gtrsim 10^{7}$ years.
This is because $\overline{\alpha_{\phi z}}$ is constrained by the cap value
$=1$ (Eq. \ref{eq:apzddp}) there.

\begin{figure} 
  \begin{center}
    \includegraphics[width=0.45\textwidth]{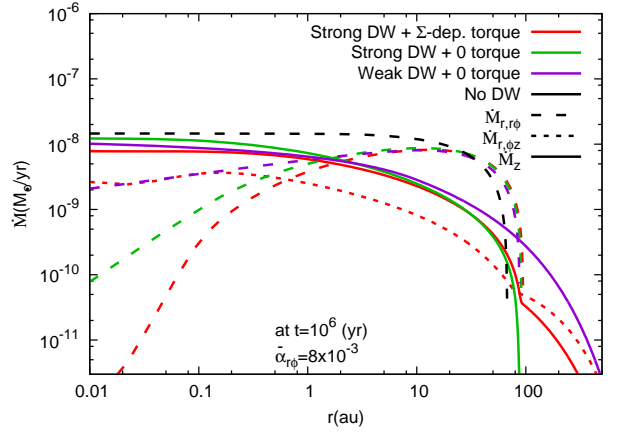}
  \end{center}
  \caption{Mass-loss rate by disc wind, $\dot{M}_z$, (solid lines)
    and mass accretion rate induced by the $r\phi$ stress, $\dot{M}_{r,r\phi}$
    (dashed lines) and by the $\phi z$ stress, $\dot{M}_{r,r\phi}$
    (dotted lines) at $t=10^6$ years of the four MRI-active cases in
    Table \ref{tab:MRIact}.
    $\dot{M}_z=0$ for the no DW case and
    $\dot{M}_{r,\phi z}=0$ for the zero-torque cases.  
    \label{fig:dtM83}
  }
\end{figure}

\begin{figure} 
  \begin{center}
    \includegraphics[width=0.45\textwidth]{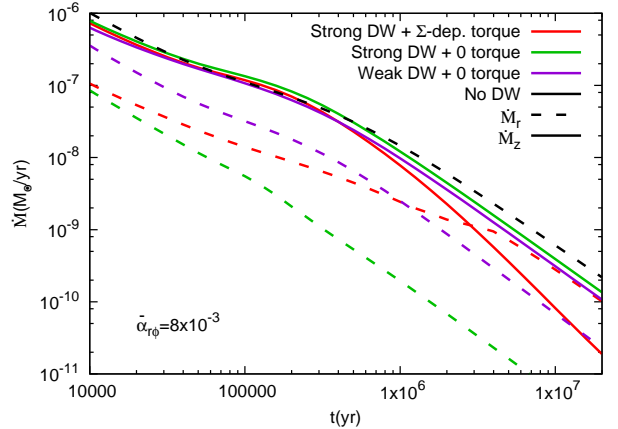}
  \end{center}
  \caption{Time evolution of $\dot{M}_z$ (solid) and
    $\dot{M}_r= \dot{M}_{r,r\phi} + \dot{M}_{r,\phi z}$ (dashed) at $r=0.0225$ au of
    the four MRI-active cases in
    Table \ref{tab:MRIact}. 
    \label{fig:t-Mdot83}
  }
\end{figure}

Figure \ref{fig:dtM83} presents the radial profile of the mass-loss rate
by disc winds (solid lines), 
\begin{equation}
  \dot{M}_z(r) = 2\pi\int_{r}^{r_{\rm out}} r dr (\rho v_{z})_{\rm w}
  =2\pi\int_{r}^{r_{\rm out}} r dr C_{\rm w}(\rho c_{\rm s})_{\rm mid}, 
  \label{eq:Mzdef}
\end{equation}
and mass accretion rate, 
\begin{equation}
  \dot{M}_r(r) = -2\pi r \Sigma v_r,
  \label{eq:Mrdef}
\end{equation}
at $t=10^6$ years.
Here, $\dot{M}_r$ can be separated into two parts,
$\dot{M}_r = \dot{M}_{r,r\phi}+\dot{M}_{r, \phi z}$, following
Eq. (\ref{eq:rdmsfx}) with help of Eqs. (\ref{eq:rphistress}) and
(\ref{eq:phizstress}) \citep[see also][]{sim13}:
Mass accretion induced by the $r\phi$ stress (dashed lines), 
\begin{equation}
  \dot{M}_{r,r\phi}(r) = - \frac{4\pi}{r\Omega}\frac{\partial}{\partial r}
  (r^2\Sigma\overline{\alpha_{r\phi}}c_{\rm s}^2), 
  \label{eq:Mrrpdef}
\end{equation}
and that by the $\phi z$ stress (dotted line)
\begin{equation}
  \dot{M}_{r,\phi z}(r) = - \frac{4\pi}{\Omega}r\overline{\alpha_{\phi z}}(\rho
  c_{\rm s}^2)_{\rm mid}.
  \label{eq:Mrpzdef}
\end{equation}
We note that $\dot{M}_z(r)$ in our definition is the total mass loss
outside $r$, while the
disc wind mass loss at $r$ is sometimes defined as the mass lost inside
$r$ \citep[e.g.,][]{owe11,bai16a,bai16b}.
We chose our definition to show how $\dot{M_r}$ is converted
into $\dot{M}_z$ as mass accretes inward.

The no DW case shows a spatially uniform accretion rate,
$\dot{M}_{r,r\phi}=1.5 \times 10^{-8}M_{\odot}$yr$^{-1}$ in $r<10$ au
(black dashed line).  When disc winds are taken into account,
the mass accretion rate decreases with decreasing $r$ as the mass is lost
by the disc winds.
When we evaluate $\dot{M}$ at $r=0.0225$ au ($\approx 4.8 R_{\odot}$), which
is one grid point outside $r_{\rm in}=0.01$ au and approximately twice 
the radius of typical T Tauri stars, $\dot{M}_{r,r\phi}$
is reduced to $2.5\times 10^{-9}M_{\odot}$yr$^{-1}$ in the weak DW
case (purple dashed line).
Instead, the mass is largely lost by the disc winds,
$\dot{M}_z=1.0\times 10^{-8} M_{\odot}$yr$^{-1}$ at $r=0.0225$ au
(purple solid line). This situation is more drastic in the strong DW + zero-torque case, and $\dot{M}_z\approx 100 \dot{M}_{r,r\phi}$ (green lines)
at $r=0.0225$ au. 
We note that $\dot{M}$ might have to be evaluated at a slightly
outer location when the inner disc is truncated by the magnetosphere of
the central star \citep[e.g.,][]{shu94,hir97,dyd15}; in this case,
$\dot{M}_r$ is not as small as the above evaluated values. 

The strong DW + $\Sigma$-dependent torque case (red lines)
gives very small $\dot{M}_{r,r\phi}$ at $r=0.0225$ au
because $\Sigma$ is small there (Fig. \ref{fig:profile83}). On the other hand,
the accretion by the $\phi z$ stress is non-zero only in this case of
the four cases displayed in Fig. \ref{fig:dtM83},
and $\dot{M}_{r,\phi z}$ is still kept $=2.4\times 10^{-9}M_{\odot}$yr$^{-1}$
at $r=0.0225$ au 
because $\overline{\alpha_{\phi z}}$ increases to $\approx 0.1$ in the inner
region; the disc is in a wind-driven accretion phase. 

Figure \ref{fig:t-Mdot83} compares the time evolutions of
$\dot{M}_z$ (solid) and $\dot{M}_r=\dot{M}_{r,r\phi}+\dot{M}_{r, \phi z}$
(dashed) at $r=0.0225$ au of these four cases.
The obtained $t-\dot{M}_r$ trends can be directly compared to the observed
distribution in the $t-\dot{M}_r$ plane \citep{gul98,har98,ric10,man16}.
Although $\dot{M}_r$ of the strong DW + zero-torque case
is smaller than the observed
lower edge ($\dot{M}_r\sim 10^{-9}M_{\odot}$yr$^{-1}$ at $t=10^6$ years),
$\dot{M}_r$ of the other three cases are well inside the observed range. 




\subsection{MRI-inactive cases}
\label{sec:MRIinactive}

\begin{table*} 
  \begin{center}
  \caption{Parameters for MRI-inactive cases
    \label{tab:MRIina}
  }
  \begin{tabular}{|c||c|c|c|c|}
    \hline
    Case & $\overline{\alpha_{r\phi}}$ & $C_{\rm w,0}$ & $\overline{\alpha_{\phi z}}$ & Energetics\\
    \hline
    \hline
    Weak DW + $\Sigma$-dependent torque & $8\times 10^{-5}$ & $10^{-5}$ &
    $10^{-5}(\Sigma/\Sigma_{\rm int})^{-0.66}$
    & Eqs.(\ref{eq:RadlossCw}) and (\ref{eq:RadlossFrad}) with
    $\epsilon_{\rm rad}=0.9$\\
    Strong DW + $\Sigma$-dependent torque & $8\times 10^{-5}$ & $10^{-5}$ &
    $10^{-5}(\Sigma/\Sigma_{\rm int})^{-0.66}$
    & Eqs.(\ref{eq:DWlossCw}) and (\ref{eq:DWlossFrad})\\
    Strong DW + constant torque & $8\times 10^{-5}$ & $10^{-5}$ & $10^{-4}$
    & Eqs.(\ref{eq:DWlossCw}) and (\ref{eq:DWlossFrad}) \\
    Strong DW + zero-torque & $8\times 10^{-5}$ & $10^{-5}$ & 0
    & Eqs.(\ref{eq:DWlossCw}) and (\ref{eq:DWlossFrad})\\
    \hline
  \end{tabular}
  \end{center}
\end{table*}

\begin{figure} 
  \begin{center}
    \includegraphics[width=0.46\textwidth]{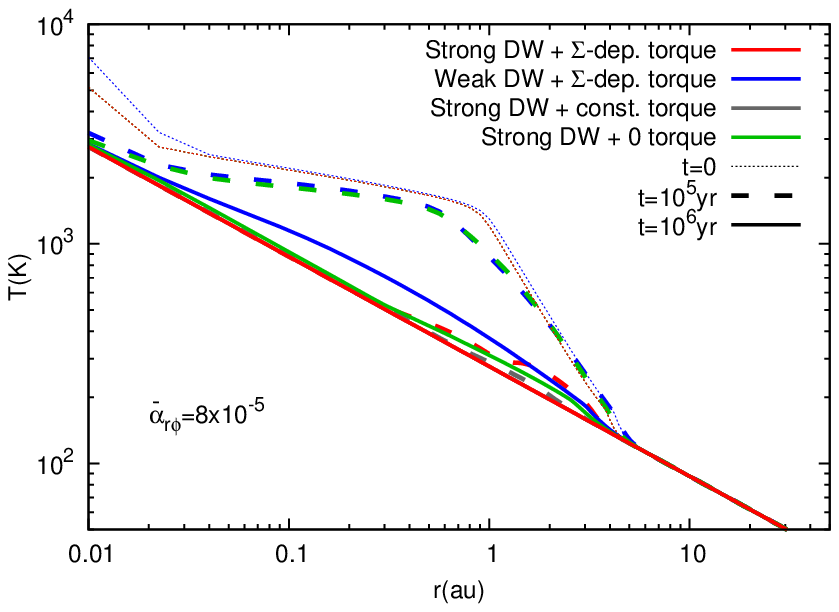}\\
    \includegraphics[width=0.47\textwidth]{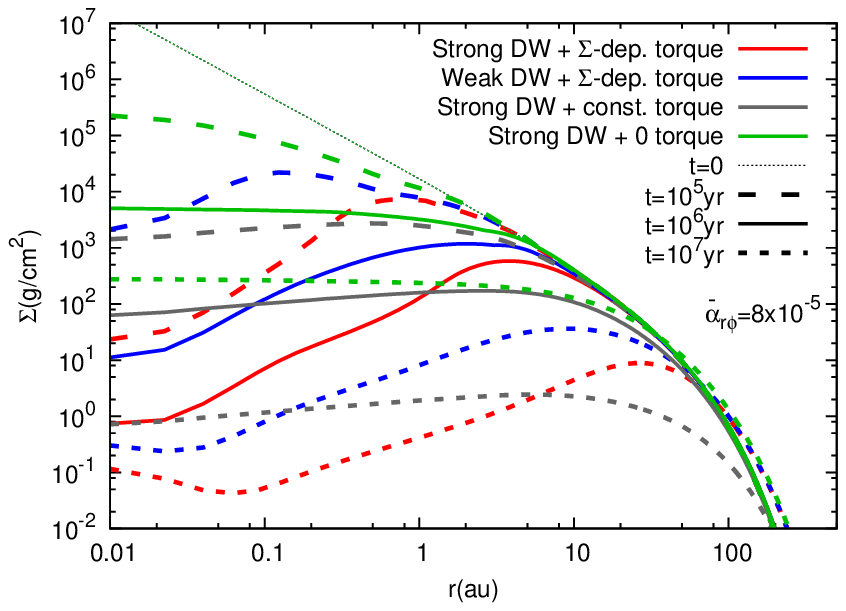}
  \end{center}
  \caption{Same as Fig. \ref{fig:profile83} but for four MRI-inactive
    cases with $\overline{\alpha_{r\phi}}=8\times 10^{-5}$. 
    The four cases are
    (i) strong DW + $\Sigma$-dependent torque (red), (ii) weak DW + $\Sigma$-dependent
    torque (blue), (iii) strong DW + constant torque (grey), and (iv) strong
    DW + zero-torque (green), summarized in Table
    \ref{tab:MRIina}. The initial temperatures of the three strong
    DW cases (red, grey, and green dotted lines) are the same and the
    red and grey solid lines at $t=10^6$ years overlap at
    $T=T_{\rm req}$ (Eq. \ref{eq:treq}). 
    \label{fig:profile85}
  }
\end{figure}

We present results of four MRI-inactive cases, which
are summarized in Table \ref{tab:MRIina}. We focus on effects of the wind
torque on the evolution of PPDs in this subsection.
Figure \ref{fig:profile85} compares radial profiles of $T$ and $\Sigma$.
The temperatures (top panel) of these cases are systematically lower
than the temperatures of the MRI-active cases (the top panel of Fig.
\ref{fig:profile83}) because smaller $\overline{\alpha_{r\phi}}$ gives
smaller $F_{\rm rad}$ (Eqs. \ref{eq:DWlossFrad} and \ref{eq:RadlossFrad})
and accordingly lower $T_{\rm vis}$ (Eq. \ref{eq:visT}). 

\begin{figure} 
  \begin{center}
    \includegraphics[width=0.45\textwidth]{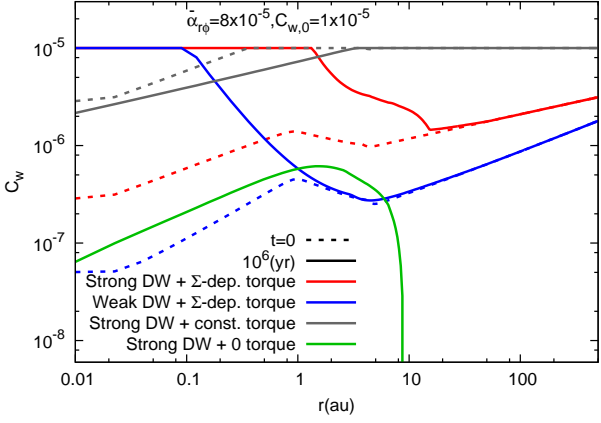}
  \end{center}
  \caption{Same as Fig. \ref{fig:Cw83} but for the MRI-inactive cases
    of Table \ref{tab:MRIina}. 
    \label{fig:Cw85}
  }
\end{figure}

Smaller $\overline{\alpha_{r\phi}}$ also leads to slower evolution;
when the MRI-active and MRI-inactive cases are compared, which adopt
the same strong DW + zero-torque parameters
(green lines in Figs. \ref{fig:profile83} and \ref{fig:profile85}),
the decrease of $\Sigma$ is much slower in the MRI-inactive case.
This is first because the accretion itself is slower owing to the smaller
$\overline{\alpha_{r\phi}}$ and second because the disc
wind mass flux is strongly constrained by the energetics to give smaller
$C_{\rm w}$ (Fig. \ref{fig:Cw85} in comparison to Fig. \ref{fig:Cw83}). 

\begin{figure} 
  \begin{center}
   \includegraphics[width=0.45\textwidth]{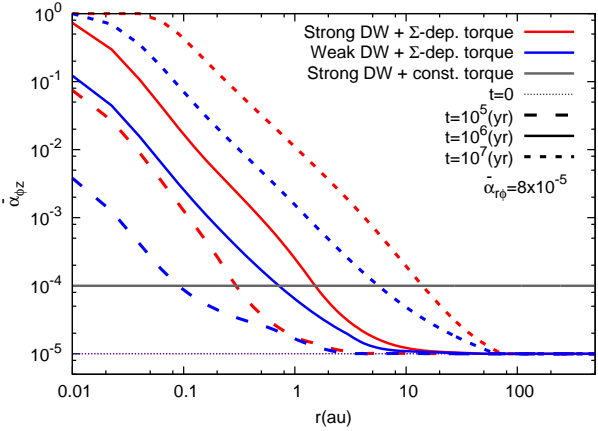}
  \end{center}
  \caption{Comparison of $\overline{\alpha_{\phi z}}$ at $t=0$ (dotted lines),
    $10^5$ (long dashed lines), $10^6$ (solid lines), and
    $10^7$ (short dashed lines) years of the non-zero-torque cases
    of Table \ref{tab:MRIina}.
    \label{fig:atrq85}
  }
\end{figure}

The evolution of $\Sigma$ is largely affected by non-zero wind torque
$\overline{\alpha_{\phi z}}$, because its effect is relatively important
for lower turbulent viscosity, $\overline{\alpha_{r\phi}}$.
The addition of the spatially constant $\overline{\alpha_{\phi z}}=10^{-4}$
(constant torque, grey lines) greatly reduces $\Sigma$.  
The two $\Sigma$-dependent torque cases
(red and blue lines) give positive slopes
of $\Sigma$ in the inner region, which we explain below. 

Figure \ref{fig:atrq85} presents the time evolution of 
$\overline{\alpha_{\phi z}}$ for the $\Sigma$-dependent torque cases.
$\overline{\alpha_{\phi z}}$ increases with time
from the inside to the outside as $\Sigma$ decreases in an inside-out manner.
As a result, the disc wind mass flux, $C_{\rm w}$, is not constrained by
the energetics (Eqs. \ref{eq:DWlossCw} and \ref{eq:RadlossCw}) but
is chosen to be the constant $C_{\rm w,0}(=10^{-5})$ in Eq. (\ref{eq:Cwe,0})
(Fig. \ref{fig:Cw85}), which leads to the inside-out dispersal of the gas.
In addition, the accretion is faster for smaller $r$ because
$\overline{\alpha_{\phi z}}$ is larger for smaller $r$.
The positive slopes of $\Sigma$ can be explained by the combination of these
effects.

Although 
we assumed spatially uniform $\overline{\alpha_{r\phi}}$
and $C_{\rm w,0}$, they are also anticipated to depend on the strength of
net vertical magnetic field, $B_z^2 /8\pi(\rho c_{\rm s}^2)_{\rm mid}$.
In this case, $\overline{\alpha_{r\phi}}$ and $C_{\rm w,0}$
could inversely correlate with $\Sigma$ \citep{suz10}, which additionally
enhances the positive slopes of the surface densities.

\begin{figure} 
  \begin{center}
    \includegraphics[width=0.47\textwidth]{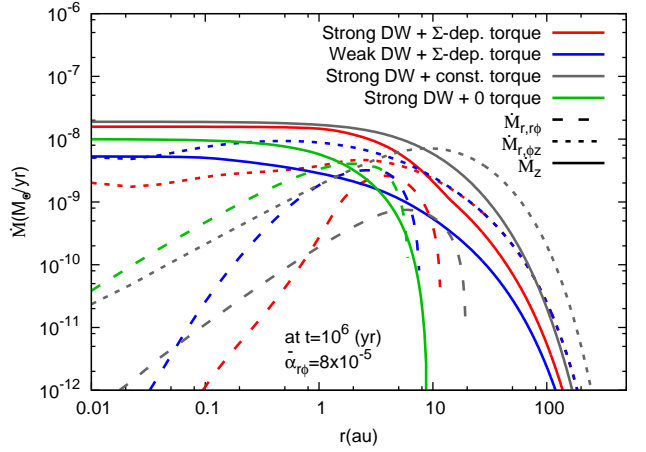}
  \end{center}
  \caption{Same as Fig. \ref{fig:dtM83} but for
    the MRI-inactive cases of Table \ref{tab:MRIina}
    \label{fig:dtM85}}
\end{figure}

\begin{figure} 
  \begin{center}
   \includegraphics[width=0.45\textwidth]{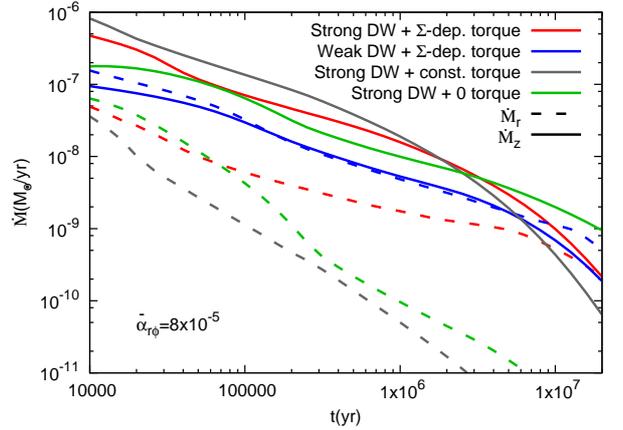}
  \end{center}
  \caption{Same as Fig. \ref{fig:t-Mdot83} but for
    the MRI-inactive cases of Table \ref{tab:MRIina}. 
    \label{fig:t-Mdot85}
  }
\end{figure}

Figure \ref{fig:dtM85} compares $\dot{M}_z$ (Eq. \ref{eq:Mzdef}; solid),
$\dot{M}_{r,r\phi}$ (Eq. \ref{eq:Mrrpdef}; dashed),
and $\dot{M}_{r,\phi z}$ (Eq. \ref{eq:Mrpzdef}; dotted) of the MRI-inactive
four cases at $t=10^6$ years. In the zero-torque case (green lines)
the mass is dominantly lost by the disc winds,
$\dot{M}_z\approx 100 \dot{M}_{r,r\phi}$ at $r=0.0225$ au.
In the constant torque case (grey lines)
the mass accretion is mainly driven by the $\phi z$ stress, 
the total accretion rate is
also largely dominated by the mass loss by the disc winds.
Adopting the $\Sigma$-dependent torque condition changes the situation;
the mass accretion is driven by the $\phi z$ stress, and the accretion rate
is well above $10^{-9}M_{\odot}$yr$^{-1}$, that is, the weak DW case gives
$\dot{M}_{r,\phi z}\approx \dot{M}_{z}\approx 5\times 10^{-9}M_{\odot}$yr$^{-1}$
at $r=0.0225$ au.

Figure \ref{fig:t-Mdot85} shows the time evolution of
$\dot{M}_z$ (solid) and $\dot{M}_r=\dot{M}_{r,r\phi}+\dot{M}_{r, \phi z}$
(dashed) at $r=0.0225$ au of the same four cases.
$\dot{M}_r(0.0225{\rm au})$ of the cases
of zero or constant torque (green and grey dashed lines) are
smaller than the observed range of $t-\dot{M}_r$
\citep{gul98,har98,ric10,man16}. On the other hand,
$\dot{M}_r(0.0225{\rm au})$'s of the $\Sigma$-dependent torque cases 
(red and blue dashed lines) are consistent with the observed $t-\dot{M}_r$.
Although the mass accretion rate of
the strong DW case is 
lower than the wind mass-loss rate (red lines),
it is not so small;
$\dot{M}_r(0.0225{\rm au})=6.0\times 10^{-9}M_{\odot}$yr$^{-1}$ at $t=10^5$ years,
$1.7\times 10^{-9}M_{\odot}$yr$^{-1}$ at $10^6$ years,
and $5.2\times10^{-10}M_{\odot}$yr$^{-1}$ at $10^7$ years.

\section{Discussion}
\subsection{Uncertainties}
\label{sec:unc}
Our model has the three free parameters, $\overline{\alpha_{r\phi}}$,
$C_{\rm w,0}$, and $\overline{\alpha_{\phi z}}$. Since these parameters are not
yet tightly constrained by observations or theoretical calculations,
we calculated the evolution of PPDs in the wide ranges of the
parameters to test various possibilities
(Sect. \ref{sec:res}).
Uncertainties of the three parameters is largely attributed to the uncertainty
of the initial distribution and to the evolution of the poloidal magnetic flux
because these three parameters 
depend on the vertical
magnetic field strength \citep{suz10,oh11,bs13b}.

 The evolution of poloidal magnetic flux in accretion
 discs has been studied by a number of groups \citep{lub94,rl08,go12,si14}
 and has recently been specifically applied to PPDs \citep{oku14,go14,to14}.
Accreting gas drags the vertical
magnetic field inward, while the vertical field also possibly diffuses outward
by magnetic diffusivity, which consists of both effective turbulent
resistivity and non-ideal MHD effects (Sect. \ref{sec:prm}). 
The radial motion of the vertical magnetic flux is determined by the balance
between these inward dragging and outward diffusion. 
The direction of the magnetic flux itself is still uncertain,
which depends on the initial
configuration of the poloidal magnetic field, in addition to the combination of
accretion and magnetic diffusion.

One future possibility is that we finally obtain a universal tendency for
the time evolution of vertical magnetic fields.
In this case, we can constrain our free parameters, and evolutions of
surface densities will not show a variety but converge to a unified trend.   
On the other hand, if the evolution of the poloidal magnetic flux
is different in different PPDs, depending on physical circumstances, such as
initial magnetic flux and disc mass, and stellar irradiation, which controls
the non-ideal MHD effects through the ionization, 
then the evolutions of surface densities are also different 
in different PPDs as shown so far, which should lead to a wide variety of
the subsequent planet formation processes and final exoplanet systems.

At present, the unified picture of the evolution of the poloidal magnetic
field is not well understood at all, and therefore it is worth pursuing
various possibilities.
Our calculations took the effect of the evolution of the vertical magnetic
field in the wind torque into account; the two cases of constant
$\overline{\alpha_{\phi z}}$ and $\Sigma$-dependent $\overline{\alpha_{\phi z}}$
correspond to the case in which the magnetic energy decreases in
the same manner as the decrease of the surface density and the case
with the preserved magnetic flux, respectively.
The $\Sigma$-dependent torque cases show a runaway behavior of the gas
dispersal in an inside-out manner; once the gas is dispersed,
$\overline{\alpha_{\phi z}}$ increases, which further accelerates the dispersal
of the gas. This is the main reason why the positive slope of $\Sigma$
is produced.
Although we did not consider this effect, $\overline{\alpha_{r\phi}}$
and $C_{\rm w,0}$ depend similarly on $\Sigma$, which causes
an additional runaway dispersal of the gas \citep[][see also Subsection
  \ref{sec:MRIinactive}]{suz10}. 
The case with constant $\overline{\alpha_{\phi z}}$ even gives
the moderately positive slope (Fig. \ref{fig:profile85}).
Within the two cases we tested, 
the positive slope of $\Sigma$ on $r$ is not peculiar, but a common feature.
However, we should note that our calculations do not cover all 
the possible distributions and evolutions of the vertical magnetic field.
Therefore, it would be 
premature to conclude that the positive slope of
$\Sigma$ is a natural outcome of the accretion induced by the magnetically
driven disc wind. 
For example, when the outward diffusion of vertical magnetic field is effective
and the magnetic flux is dispersed more rapidly than the gas, the effect of
the wind torque is suppressed with time. In this case, the 
$\Sigma$ profile would maintain a normal negative slope.

We now discuss other ambiguities of the mass flux of the disc winds,
in addition to the uncertainty of the vertical magnetic field. 
At the moment, the mass flux,  $C_{\rm w,0}$, is available only from local MHD
simulations \citep[e.g.][]{si09,fro13,bs13a}. As discussed 
in Sect. \ref{sec:prm},
these local simulations may overestimate the mass flux. Although we adopted
the conservative $C_{\rm w,0}$ by reducing the simulation results by half
(see Sect. \ref{sec:prm}), it might be even lower \citep{fro13}.
We here briefly discuss how the results are affected and particularly focus
on the slope of the surface density when $C_{\rm w,0}$ is smaller.

As shown in Figs. \ref{fig:Cw83} and \ref{fig:Cw85}$, C_{\rm w}$ is
already constrained by the energetics. In most cases except for the
MRI-inactive cases with $\Sigma$-dependent torque, the energetics constraint
already suppresses $C_{\rm w}$ in the inner region. 
Therefore, adopting a smaller $C_{\rm w,0}$ does not affect $C_{\rm w}$ in the
inner region but reduces $C_{\rm w}$ in the outer region,
which suppresses the gas dispersal there.
Hence, the slope of $\Sigma$ would be more positive in these cases.
On the other hand, in the MRI-inactive cases with $\Sigma$-dependent torque,
the energetics constraint suppresses $C_{\rm w}$ at the relatively outer
location, $r\sim 10$ au. In these cases, a smaller $C_{\rm w,0}$ reduces
$C_{\rm w}$ in the inner region. As a result, the obtained large positive
$\Sigma$ slopes in these cases (Fig. \ref{fig:Cw85}) would be
reduced to moderately positive ones. 

When we determined the mass flux of the disc winds, we applied the
energetics constraint from the gravitational accretion without external
heating or momentum inputs (Sect. \ref{sec:Cw}; Eq. \ref{eq:Cwe,0}).
This treatment is expected to give a reasonable constraint at the early
phase when viscous heating dominates the radiative heating or other
effects from the central star. However, at the later time this is not
the case because the surface density decreases and the viscous heating
becomes relatively unimportant.
Effects of external heating or momentum inputs need to be considered.
They weaken the energetics constraint to give a larger $C_{\rm w}$
in the region with $C_{\rm w,e} < C_{\rm w,0}$ (see Sect. \ref{sec:swpe}).

\subsection{Radial drift of pebbles and boulders}
\label{sec:drift}

\begin{figure} 
  \begin{center}
    \includegraphics[width=0.45\textwidth]{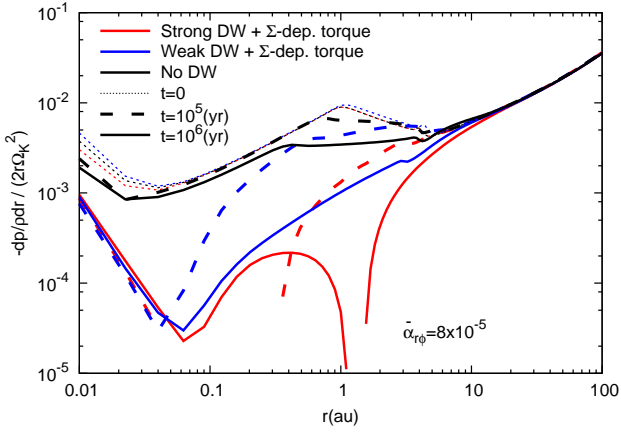}
  \end{center}
  \caption{Comparison of normalized pressure gradient force,
    $-\left(\frac{1}{\rho_{\rm mid}}\frac{\partial p_{\rm mid}}{\partial r}\right)
    /(2r\Omega^2)$, of MRI-inactive PPDs 
    at $t=0$ (dotted), $10^5$ (solid), and $10^6$ years (dashed). 
    The MRI-inactive cases with $\Sigma$-dependent torque in
    Table \ref{tab:MRIina},
    blue lines for weak DW and red lines for strong DW, which
    corresponds to the red and blue lines in Fig. \ref{fig:profile85},
    are compared to the MRI-inactive no DW case with $C_{\rm w,0}=0$ and
    $\overline{\alpha_{r\phi}}
    =8\times 10^{-5}$ (black lines).     
    \label{fig:pgrf}
  }
\end{figure}

Although calculations still include uncertainties that
mainly stem from the ambiguity of the evolution of poloidal magnetic fields,
the positive slopes of the surface densities obtained in Sect. \ref{sec:res}
are 
a possible
consequence of the evolution of PPDs with
disc winds, as discussed in Sect. \ref{sec:unc}. 
These positive slopes 
raise various interesting implications for planet formation.
In this and the next subsections, we demonstrate how the obtained $\Sigma$
profiles affect the solid component of PPDs by studying cases that show large
positive slopes of $\Sigma$. 

The first example is the radial drift of solid bodies through gas drag.
In general the rotation velocity of the gas in PPDs is slightly slower
than the local Keplerian velocity because of
the radial pressure gradient force. On the other hand,
solid particles rotate with Keplerian velocity without the support
from the gas pressure.
As a result, the solid particles feel a head wind from the gas, which causes
them to drift inward. Considering the momentum balance, solid particles with
nondimensional stopping time $\approx 1$, 
which corresponds to
a meter-sized spherical boulder at 1 au of the MMSN,  
experience the radial drift most severely
\citep{wei77,nak86}, and their drift timescale in the midplane is given by
\begin{equation}
\tau_{\rm dr,max} \approx \frac{1}{\eta \Omega_{\rm K}}, 
\end{equation}
where $\eta$ is pressure gradient force normalized by the twice of
centrifugal force,
\begin{equation}
  \eta = -\frac{1}{\rho_{\rm mid}}\frac{\partial p_{\rm mid}}{\partial r}
  \frac{1}{2r\Omega_{\rm K}^2}. 
\end{equation}
In the usual condition, $\eta\sim 10^{-3}-10^{-2}>0$, which causes solid
particles to drift inward. Smaller $\eta$ leads to slower inward drift;
if $\eta < 0$, the
direction of the drift is opposite and solid particles move outward.

Figure \ref{fig:pgrf} shows $\eta$ of the two MRI-inactive
($\overline{\alpha_{r\phi}}=8\times 10^{-5}$) cases with
$\Sigma$-dependent torque  of Table \ref{tab:MRIina}
(red and blue lines; the same as in Figs. \ref{fig:profile85}
-- \ref{fig:t-Mdot85})
in comparison to the no disc wind (no DW) case with the same 
$\overline{\alpha_{r\phi}}=8\times 10^{-5}$ (black lines).
We here derive $p_{\rm mid}$ from $\Sigma$ by  
\begin{equation}
  p_{\rm mid} = \rho_{\rm mid} c_{\rm s}^2 = \frac{\Sigma\Omega c_{\rm s}}{\sqrt{2\pi}}
\end{equation}
The no DW case shows $\eta$ remains within $10^{-3}-10^{-2}$,
which implies fast inward drift.
In contrast, $\eta$'s are considerably reduced in the $\Sigma$-dependent torque
cases. 
In particular, the red lines (strong DW case) show
negative $\eta$ in part (red lines are truncated between 0.04-0.4 au
at $t=10^5$ years and 1-2 au at $t=10^6$ years), which indicates that solid
particles move outward in this region.
As a result, the solid component will accumulate around the
outer edge of the negative $\eta$ region, which offers suitable conditions
for planet formation \citep{kob12}.
Furthermore, this location moves outward with time; the suitable site for the
planet formation also moves outward.

\subsection{Type I migration}
\label{sec:TypeImig}
\begin{figure} 
  \begin{center}
    \includegraphics[width=0.45\textwidth]{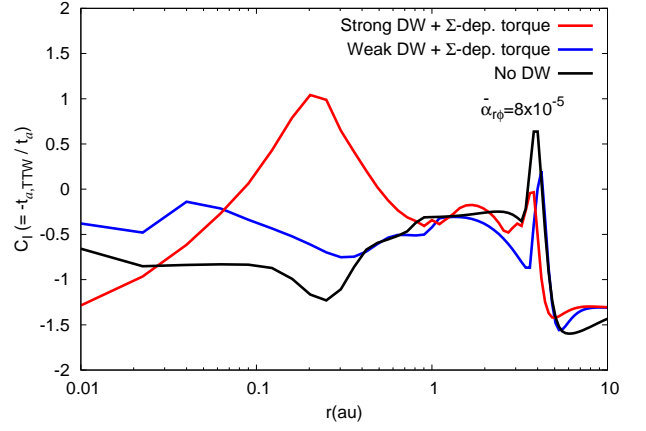}
  \end{center}
  \caption{Migration efficiency for Earth-mass planets for 
    MRI-inactive cases with $\Sigma$-dependent torque (red for strong DW
    and blue for weak DW of Table \ref{tab:MRIina}) 
    at $t=10^6$ yr  (corresponding to the solid red and blue lines in
    Fig. \ref{fig:profile85}) in comparison to the MRI-inactive no DW case
    (black).
    $C_{\rm I} > 0$ means outward migration.
    \label{fig:TypeImig}
  }
\end{figure}

Another interesting implication of the positive $\Sigma$ slopes is
that an inward migration of low-mass planets (type I migration) can be
slowed down or even reversed. The torque for type I migration can be
expressed by the sum of Lindblad and corotation torques. The
corotation torque is more sensitive to the slope of the gas surface
density and can be positive for positive slopes.

Here we estimate the migration rate of Earth-mass planets embedded in
MRI-inactive PPDs with the surface densities shown
in Fig.~\ref{fig:profile85}.
We used the formulae of \citet{paa11} to calculate 
the migration timescale, $t_a$ (see Eqs.~(8)-(16) in \citet{ogi15a}
for details of the formulae). We introduced a parameter of the
efficiency of inward type I migration, $C_{\rm I} \equiv -t_{a,{\rm
TTW}}/t_a$, where $t_{a,{\rm TTW}}$ is the migration time in a locally
isothermal disc derived by a linear analysis by \citet{tan02}.
The migration timescale is defined as $t_a \equiv
a/(-\dot{a})$; positive migration time means inward migration.

Figure \ref{fig:TypeImig} shows the migration efficiency for the $\Sigma$-dependent
torque case (the red and blue curves in Fig.~\ref{fig:profile85}) at
$t = 10^6$yr in comparison to the no DW case (black line). The migration rate
depends on the planetary mass and the orbital eccentricity; Earth-mass
planets with zero eccentricity were considered here. The blue curve
shows that the type I migration is slowed down inside a few au by
several factors from $t_{a,{\rm TTW}}$. The migration is even reversed
(outward migration) between 0.1-0.5 au in the red curve (strong DW case).
Thus the disc wind would also play important roles in the late stage of
planet formation.

\subsection{Comparison to previous work}
\label{sec:comp}
Recently, \citet{bai16b} also presented a global evolution model for PPDs with
magnetically driven disc winds.
However, none of the cases in his model calculations resulted in a surface density with
a drastic positive slope relative to $r$ as some of our cases have shown.
The two main differences between his setup and ours is the mass-loss rate
by the disc wind and the evolution of the vertical magnetic field.

Our calculations, which started from a relatively massive initial disc
($M_{\rm disc,int}=0.11M_{\odot}$) to study the evolution from the early stage,
neglected the heating by the irradiation from a central star but considered
viscous heating, and the mass-loss rate was constrained by the global
energetics of the viscous accretion.
In contrast, the initial disc mass
adopted by \citet{bai16b} is lower, $=0.035M_{\odot}$, to focus on the later
stage of the evolution, and the location of the wind base in the inner
region $r\lesssim 10-30$ au is determined from heating by far-ultraviolet
(FUV hereafter) irradiation from a central star.
Here, the penetration depth of the FUV was assumed to be spatially constant.
Since the surface density decreases with $r$, the penetration depth normalized
by the scale height is deeper for larger $r$. Therefore, the mass loss
by the disc wind affects the depletion of the gas at outer locations more
severely than in our model setting, and consequently a positive slope of
$\Sigma$ was not obtained in the results of \citet{bai16b}.

As for the evolution of the vertical magnetic field, \citet{bai16b}
considered two cases: in the first case the total magnetic flux is preserved
with time, and in the second case it decreases in the same manner
as the total mass. 
In both cases, the plasma $\beta = (B_z^2/8\pi(\rho
c_{\rm s}^2)_{\rm mid})^{-1}$ at the midplane was assumed to be spatially uniform.
Even in the first case, the vertical magnetic field was redistributed to
follow the density profile \citep{arm13}.
This spatially uniform $\beta$ was also adopted in our constant torque
setting. 
In contrast, our $\Sigma$-dependent torque assumed the preserved vertical
magnetic field at each location, which led to a runaway inside-out
dispersal and produced a large positive slope of $\Sigma$
(Sect. \ref{sec:res}), compared to the above-mentioned cases with the
spatially uniform $\beta$.


\subsection{Stellar wind and photoevaporation}
\label{sec:swpe}
We did not take the effects of a central star into account 
except to determine the radiative equilibrium temperature, $T_{\rm req}$
(Eq. \ref{eq:treq}). However, the stellar wind and irradiation affect
the evolution of PPDs. 

In our calculations, the mass flux of the disc wind is $C_{\rm w,e}$ constrained
by the energetics of accretion, and 
it can be smaller than $C_{\rm w,0}$ determined by the mass loading expected
from the local MHD simulations. When this is the case, gaseous clouds are
lifted up by vertical upflows but cannot stream out to large $z$; they float
in the disc atmosphere or return to the disc because they are bound
by the gravity of the central star. The stellar wind from the central star
would change this situation. 

The mass flux of the stellar wind from pre-main sequence stars is much higher,
by an order of 4 -- 6, than that of the current solar wind partly because of
the energy supply from accretion \citep{hir97,mp05,cra09}.
Even after the accretion terminates, the mass flux of the stellar wind is
expected to be still high because of the high magnetic activity \citep{wood05,cs11,suz13}. 
The strong stellar wind would blow away the clouds that are lifted up by the disc winds
\citep[see][for the energetics]{suz10}.
In the framework of our model, the contribution from the stellar wind would
increase $C_{\rm w,e}$ in Eqs. (\ref{eq:Cwcn1}) and (\ref{eq:Cwcn2}),
in the small $r$ region. The increase of $C_{\rm w}$ in the inner region
reduces $\Sigma$ there, which also produces a larger positive slope of $\Sigma$.

In this discussion, we neglected the roles of global magnetic fields that are 
rooted in the central star and in the PPD.
When the field strength is strong enough,
the stellar wind region and the disc wind region are separated by a boundary
layer formed by magnetospheric ejections \citep{zf13}. In this case
the stellar winds will not contribute to driving the disc winds.
It depends on the relative strength of the magnetic energy to the sum of
the dynamic pressure and the gas pressure whether the interaction between
the stellar winds and the disc winds is efficient. 
When the magnetic energy is weaker, the interaction is stronger, and vice versa.

Photoevaporation by irradiation from the central star or neighbouring
stars has been extensively studied as a viable source for dispersing PPDs
\citep[e.g.,][]{shu93,hol00,adm04}.
The mass-loss rate by the photoevaporation, which depends on the flux
in different spectral ranges, FUV, extreme UV, and X-rays,
yields a wide variety of
$\sim 10^{-10}-10^{-8}M_{\odot}$yr$^{-1}$ \citep{alx06,erc08,gor09,owe10,tan13}.
After the mass accretion rate or the mass-loss rate by the disc wind decreases
below this level, the photoevaporation would quickly disperse PPDs
\citep[e.g.][]{arm11}; our results would be affected at the late stage of
the evolution. 

However, we expect that the evolution of the $\Sigma$ profile of a
photoevaporating PPD is qualitatively different from our results with the
magnetically driven disc wind because the photoevaporation mostly affects
the disc dispersal in the outer region where the sound speed of the heated
gas exceeds the local escape velocity from the central star. 
Although the photoevaporation could create an inner hole by the
combination with the viscous accretion, the local slope of $\Sigma$ remains
negative except at the inner edge of the hole \citep[e.g.][]{alx06,owe11}. 
This is in clear contrast to the evolution with the magnetically driven
disc wind. 

\section{Summary}
We have studied the global evolution of PPDs by considering viscous heating
and magnetically driven disc winds. We constructed a global model from
fundamental MHD equations for the time-evolution of PPDs. One of the key
features of our model is that the mass-loss rate by the disc wind is
derived from both the local MHD shearing box simulations and the global
energetics of the gravitational accretion.
Our model has three dimensionless parameters, which are turbulence
viscosity, $\overline{\alpha_{r\phi}}$, disc wind mass flux,
$C_{\rm w}$, and disc wind torque, $\overline{\alpha_{\phi z}}$, and
these three parameters are 
constrained by the above-mentioned global energetics. 
We performed model calculations in a wide parameter range
to cover both MRI-active PPDs and MRI-inactive PPDs with dead zones. 

We started our calculations from the relatively massive disc,
$M_{\rm disc.int}=0.11M_{\odot}$. Initially, the viscous heating dominantly
determines the temperature in the inner region $<10$ au; for instance,
$T\simeq 1500$ K at 1 au, which is much higher than the temperature
estimated from the radiative equilibrium.
As the surface density decreases with time, the temperature approaches
the radiative equilibrium temperature.
In the cases that consider the disc wind mass loss, the gas in the inner
region is rapidly dispersed before $10^6$ years, and the viscous
heating is negligible in determining the temperature after $t\gtrsim 10^6$
years, whereas in the no disc wind cases the viscous heating is not
negligible even up to several $10^6$ years.

The mass accretion rates decrease with time as the surface
densities decrease, regardless of whether the accretion is induced by
turbulent viscosity or wind torque. The obtained accretion rates
are consistent with observed accretion rates for a wide range of the
adopted parameters. 

The three free parameters, $\overline{\alpha_{r\phi}}$,
$C_{\rm w,0}$, and $\overline{\alpha_{\phi z}}$ still contain ambiguities,
arising mainly from the uncertainty of the evolution of vertical magnetic
fields. We have pursued various possibilities by testing different
combinations of these parameters. 
The detailed profiles of the temperatures and the surface densities show
a wide variety.
Since physical properties of a PPD affect planet formation
processes that take place in the disc \citep[e.g.,][]{kob16},
the obtained variety of our PPD calculations would be a source of
the observed variety of exoplanet systems \citep[e.g.][]{and12}.

The wind-driven accretion can promote an increase in disc
  surface density with $r$ in the inner region; this is the case in our
  calculations for MRI-inactive PPDs when the distribution of the vertical
  magnetic flux is preserved with time evolution 
(Sect. \ref{sec:MRIinactive}).
This large positive slope of the surface density suppresses or reverses
the inward drift of pebble- or boulder-sized solids through gas drag (Sect.
\ref{sec:drift}) and the inward migration of protoplanets (Sect.
\ref{sec:TypeImig}), which is a favourable condition for the planet
formation. 

T.K.S. is supported by the Astrobiology Center Project of National Institute
of Natural Sciences (NINS) (Grant Number AB271020, AB281018).
A.M., A.C. and T.G. acknowledge support by the French ANR, project number
ANR-13--13-BS05-0003-01  project MOJO (Modeling the Origin of JOvian planets).
T.K.S. thanks Hiroshi Kobayashi and Shinsuke Takasao for fruitful
discussions. The authors thank the referee for many valuable comments.

\nocite{*}
\bibliography{paper_dwtrq201615aa}

\begin{thebibliography}{97}
\expandafter\ifx\csname natexlab\endcsname\relax\def\natexlab#1{#1}\fi

\bibitem[{{Adams} {et~al.}(2004){Adams}, {Hollenbach}, {Laughlin}, \&
  {Gorti}}]{adm04}
{Adams}, F.~C., {Hollenbach}, D., {Laughlin}, G., \& {Gorti}, U. 2004, \apj,
  611, 360

\bibitem[{{Alexander} {et~al.}(2006){Alexander}, {Clarke}, \&
  {Pringle}}]{alx06}
{Alexander}, R.~D., {Clarke}, C.~J., \& {Pringle}, J.~E. 2006, \mnras, 369, 229

\bibitem[{{Armitage}(2011)}]{arm11}
{Armitage}, P.~J. 2011, \araa, 49, 195

\bibitem[{{Armitage} {et~al.}(2013){Armitage}, {Simon}, \& {Martin}}]{arm13}
{Armitage}, P.~J., {Simon}, J.~B., \& {Martin}, R.~G. 2013, \apjl, 778, L14

\bibitem[{{Bai}(2013)}]{bai13}
{Bai}, X.-N. 2013, \apj, 772, 96

\bibitem[{{Bai}(2016)}]{bai16b}
{Bai}, X.-N. 2016, \apj, 821, 80

\bibitem[{{Bai} \& {Stone}(2013{\natexlab{a}})}]{bs13a}
{Bai}, X.-N. \& {Stone}, J.~M. 2013{\natexlab{a}}, \apj, 767, 30

\bibitem[{{Bai} \& {Stone}(2013{\natexlab{b}})}]{bs13b}
{Bai}, X.-N. \& {Stone}, J.~M. 2013{\natexlab{b}}, \apj, 769, 76

\bibitem[{{Bai} {et~al.}(2016){Bai}, {Ye}, {Goodman}, \& {Yuan}}]{bai16a}
{Bai}, X.-N., {Ye}, J., {Goodman}, J., \& {Yuan}, F. 2016, \apj, 818, 152

\bibitem[{{Bailli{\'e}} {et~al.}(2015){Bailli{\'e}}, {Charnoz}, \&
  {Pantin}}]{bail15}
{Bailli{\'e}}, K., {Charnoz}, S., \& {Pantin}, E. 2015, \aap, 577, A65

\bibitem[{{Balbus} \& {Hawley}(1991)}]{bh91}
{Balbus}, S.~A. \& {Hawley}, J.~F. 1991, \apj, 376, 214

\bibitem[{{Balbus} \& {Hawley}(1998)}]{bh98}
{Balbus}, S.~A. \& {Hawley}, J.~F. 1998, Reviews of Modern Physics, 70, 1

\bibitem[{{Bitsch} {et~al.}(2015){Bitsch}, {Johansen}, {Lambrechts}, \&
  {Morbidelli}}]{bit15}
{Bitsch}, B., {Johansen}, A., {Lambrechts}, M., \& {Morbidelli}, A. 2015, \aap,
  575, A28

\bibitem[{{Blandford} \& {Payne}(1982)}]{bp82}
{Blandford}, R.~D. \& {Payne}, D.~G. 1982, \mnras, 199, 883

\bibitem[{{Chandrasekhar}(1961)}]{cha61}
{Chandrasekhar}, S. 1961, {Hydrodynamic and hydromagnetic stability} (Oxford:
  Clarendon)

\bibitem[{{Chiang} \& {Youdin}(2010)}]{cy10}
{Chiang}, E. \& {Youdin}, A.~N. 2010, Annual Review of Earth and Planetary
  Sciences, 38, 493

\bibitem[{{Chiang} \& {Goldreich}(1997)}]{cg97}
{Chiang}, E.~I. \& {Goldreich}, P. 1997, \apj, 490, 368

\bibitem[{{Cranmer}(2009)}]{cra09}
{Cranmer}, S.~R. 2009, \apj, 706, 824

\bibitem[{{Cranmer} \& {Saar}(2011)}]{cs11}
{Cranmer}, S.~R. \& {Saar}, S.~H. 2011, \apj, 741, 54

\bibitem[{{Davis}(2005)}]{dav05}
{Davis}, S.~S. 2005, \apj, 620, 994

\bibitem[{{Dullemond} {et~al.}(2002){Dullemond}, {van Zadelhoff}, \&
  {Natta}}]{dul02}
{Dullemond}, C.~P., {van Zadelhoff}, G.~J., \& {Natta}, A. 2002, \aap, 389, 464

\bibitem[{{Dyda} {et~al.}(2015){Dyda}, {Lovelace}, {Ustyugova}, {Lii},
  {Romanova}, \& {Koldoba}}]{dyd15}
{Dyda}, S., {Lovelace}, R.~V.~E., {Ustyugova}, G.~V., {et~al.} 2015, \mnras,
  450, 481

\bibitem[{{Dzyurkevich} {et~al.}(2013){Dzyurkevich}, {Turner}, {Henning}, \&
  {Kley}}]{dzy13}
{Dzyurkevich}, N., {Turner}, N.~J., {Henning}, T., \& {Kley}, W. 2013, \apj,
  765, 114

\bibitem[{{Ercolano} {et~al.}(2008){Ercolano}, {Drake}, {Raymond}, \&
  {Clarke}}]{erc08}
{Ercolano}, B., {Drake}, J.~J., {Raymond}, J.~C., \& {Clarke}, C.~C. 2008,
  \apj, 688, 398

\bibitem[{{Ferreira} {et~al.}(2006){Ferreira}, {Dougados}, \& {Cabrit}}]{fer06}
{Ferreira}, J., {Dougados}, C., \& {Cabrit}, S. 2006, \aap, 453, 785

\bibitem[{{Flock} {et~al.}(2012){Flock}, {Henning}, \& {Klahr}}]{flo12}
{Flock}, M., {Henning}, T., \& {Klahr}, H. 2012, \apj, 761, 95

\bibitem[{{Fromang} {et~al.}(2013){Fromang}, {Latter}, {Lesur}, \&
  {Ogilvie}}]{fro13}
{Fromang}, S., {Latter}, H., {Lesur}, G., \& {Ogilvie}, G.~I. 2013, \aap, 552,
  A71

\bibitem[{{Gammie}(1996)}]{gam96}
{Gammie}, C.~F. 1996, \apj, 457, 355

\bibitem[{{Garaud} \& {Lin}(2007)}]{gl07}
{Garaud}, P. \& {Lin}, D.~N.~C. 2007, \apj, 654, 606

\bibitem[{{Gorti} \& {Hollenbach}(2009)}]{gor09}
{Gorti}, U. \& {Hollenbach}, D. 2009, \apj, 690, 1539

\bibitem[{{Gressel} {et~al.}(2015){Gressel}, {Turner}, {Nelson}, \&
  {McNally}}]{gre15}
{Gressel}, O., {Turner}, N.~J., {Nelson}, R.~P., \& {McNally}, C.~P. 2015,
  \apj, 801, 84

\bibitem[{{Guilet} \& {Ogilvie}(2012)}]{go12}
{Guilet}, J. \& {Ogilvie}, G.~I. 2012, \mnras, 424, 2097

\bibitem[{{Guilet} \& {Ogilvie}(2014)}]{go14}
{Guilet}, J. \& {Ogilvie}, G.~I. 2014, \mnras, 441, 852

\bibitem[{{Gullbring} {et~al.}(1998){Gullbring}, {Hartmann}, {Brice{\~n}o}, \&
  {Calvet}}]{gul98}
{Gullbring}, E., {Hartmann}, L., {Brice{\~n}o}, C., \& {Calvet}, N. 1998, \apj,
  492, 323

\bibitem[{{Haisch} {et~al.}(2001){Haisch}, {Lada}, \& {Lada}}]{hai01}
{Haisch}, Jr., K.~E., {Lada}, E.~A., \& {Lada}, C.~J. 2001, \apjl, 553, L153

\bibitem[{{Hartmann} {et~al.}(1998){Hartmann}, {Calvet}, {Gullbring}, \&
  {D'Alessio}}]{har98}
{Hartmann}, L., {Calvet}, N., {Gullbring}, E., \& {D'Alessio}, P. 1998, \apj,
  495, 385

\bibitem[{{Hawley} {et~al.}(2011){Hawley}, {Guan}, \& {Krolik}}]{haw11}
{Hawley}, J.~F., {Guan}, X., \& {Krolik}, J.~H. 2011, \apj, 738, 84

\bibitem[{{Hayashi}(1981)}]{hay81}
{Hayashi}, C. 1981, Progress of Theoretical Physics Supplement, 70, 35

\bibitem[{{Hayashi} {et~al.}(1985){Hayashi}, {Nakazawa}, \& {Nakagawa}}]{hay85}
{Hayashi}, C., {Nakazawa}, K., \& {Nakagawa}, Y. 1985, in Protostars and
  Planets II, ed. D.~C. {Black} \& M.~S. {Matthews}, 1100--1153

\bibitem[{{Hern{\'a}ndez} {et~al.}(2008){Hern{\'a}ndez}, {Hartmann}, {Calvet},
  {Jeffries}, {Gutermuth}, {Muzerolle}, \& {Stauffer}}]{her08}
{Hern{\'a}ndez}, J., {Hartmann}, L., {Calvet}, N., {et~al.} 2008, \apj, 686,
  1195

\bibitem[{{Hirose} \& {Turner}(2011)}]{ht11}
{Hirose}, S. \& {Turner}, N.~J. 2011, \apjl, 732, L30

\bibitem[{{Hirose} {et~al.}(1997){Hirose}, {Uchida}, {Shibata}, \&
  {Matsumoto}}]{hir97}
{Hirose}, S., {Uchida}, Y., {Shibata}, K., \& {Matsumoto}, R. 1997, \pasj, 49,
  193

\bibitem[{{Hollenbach} {et~al.}(2000){Hollenbach}, {Yorke}, \&
  {Johnstone}}]{hol00}
{Hollenbach}, D.~J., {Yorke}, H.~W., \& {Johnstone}, D. 2000, Protostars and
  Planets IV, 401

\bibitem[{{Howard} {et~al.}(2012){Howard}, {Marcy}, {Bryson}, {Jenkins},
  {Rowe}, {Batalha}, {Borucki}, {Koch}, {Dunham}, {Gautier}, {Van Cleve},
  {Cochran}, {Latham}, {Lissauer}, {Torres}, {Brown}, {Gilliland}, {Buchhave},
  {Caldwell}, {Christensen-Dalsgaard}, {Ciardi}, {Fressin}, {Haas}, {Howell},
  {Kjeldsen}, {Seager}, {Rogers}, {Sasselov}, {Steffen}, {Basri},
  {Charbonneau}, {Christiansen}, {Clarke}, {Dupree}, {Fabrycky}, {Fischer},
  {Ford}, {Fortney}, {Tarter}, {Girouard}, {Holman}, {Johnson}, {Klaus},
  {Machalek}, {Moorhead}, {Morehead}, {Ragozzine}, {Tenenbaum}, {Twicken},
  {Quinn}, {Isaacson}, {Shporer}, {Lucas}, {Walkowicz}, {Welsh}, {Boss},
  {Devore}, {Gould}, {Smith}, {Morris}, {Prsa}, {Morton}, {Still}, {Thompson},
  {Mullally}, {Endl}, \& {MacQueen}}]{and12}
{Howard}, A.~W., {Marcy}, G.~W., {Bryson}, S.~T., {et~al.} 2012, \apjs, 201, 15

\bibitem[{{Hueso} \& {Guillot}(2005)}]{hg05}
{Hueso}, R. \& {Guillot}, T. 2005, \aap, 442, 703

\bibitem[{{Johansen} {et~al.}(2009){Johansen}, {Youdin}, \& {Klahr}}]{joh09}
{Johansen}, A., {Youdin}, A., \& {Klahr}, H. 2009, \apj, 697, 1269

\bibitem[{{Kimura} {et~al.}(2016){Kimura}, {Kunitomo}, \& {Takahashi}}]{kim16}
{Kimura}, S.~S., {Kunitomo}, M., \& {Takahashi}, S.~Z. 2016, \mnras, 461, 2257

\bibitem[{{Kobayashi} {et~al.}(2012){Kobayashi}, {Ormel}, \& {Ida}}]{kob12}
{Kobayashi}, H., {Ormel}, C.~W., \& {Ida}, S. 2012, \apj, 756, 70

\bibitem[{{Kobayashi} {et~al.}(2016){Kobayashi}, {Tanaka}, \&
  {Okuzumi}}]{kob16}
{Kobayashi}, H., {Tanaka}, H., \& {Okuzumi}, S. 2016, \apj, 817, 105

\bibitem[{{Kusaka} {et~al.}(1970){Kusaka}, {Nakano}, \& {Hayashi}}]{kus70}
{Kusaka}, T., {Nakano}, T., \& {Hayashi}, C. 1970, Progress of Theoretical
  Physics, 44, 1580

\bibitem[{{Lesur} {et~al.}(2013){Lesur}, {Ferreira}, \& {Ogilvie}}]{les13}
{Lesur}, G., {Ferreira}, J., \& {Ogilvie}, G.~I. 2013, \aap, 550, A61

\bibitem[{{Lesur} \& {Longaretti}(2007)}]{ll07}
{Lesur}, G. \& {Longaretti}, P.-Y. 2007, \mnras, 378, 1471

\bibitem[{{Lubow} {et~al.}(1994){Lubow}, {Papaloizou}, \& {Pringle}}]{lub94}
{Lubow}, S.~H., {Papaloizou}, J.~C.~B., \& {Pringle}, J.~E. 1994, \mnras, 267,
  235

\bibitem[{{Lynden-Bell} \& {Pringle}(1974)}]{lp74}
{Lynden-Bell}, D. \& {Pringle}, J.~E. 1974, \mnras, 168, 603

\bibitem[{{Manara} {et~al.}(2016){Manara}, {Fedele}, {Herczeg}, \&
  {Teixeira}}]{man16}
{Manara}, C.~F., {Fedele}, D., {Herczeg}, G.~J., \& {Teixeira}, P.~S. 2016,
  \aap, 585, A136

\bibitem[{{Matt} \& {Pudritz}(2005)}]{mp05}
{Matt}, S. \& {Pudritz}, R.~E. 2005, \apjl, 632, L135

\bibitem[{{Miyake} {et~al.}(2016){Miyake}, {Suzuki}, \& {Inutsuka}}]{miy16}
{Miyake}, T., {Suzuki}, T.~K., \& {Inutsuka}, S.-i. 2016, \apj, 821, 3

\bibitem[{{Nakagawa} {et~al.}(1986){Nakagawa}, {Sekiya}, \& {Hayashi}}]{nak86}
{Nakagawa}, Y., {Sekiya}, M., \& {Hayashi}, C. 1986, ICARUS, 67, 375

\bibitem[{{Nakamoto} \& {Nakagawa}(1994)}]{nn94}
{Nakamoto}, T. \& {Nakagawa}, Y. 1994, \apj, 421, 640

\bibitem[{{Ogihara} {et~al.}(2015{\natexlab{a}}){Ogihara}, {Kobayashi},
  {Inutsuka}, \& {Suzuki}}]{ogi15a}
{Ogihara}, M., {Kobayashi}, H., {Inutsuka}, S.-i., \& {Suzuki}, T.~K.
  2015{\natexlab{a}}, \aap, 579, A65

\bibitem[{{Ogihara} {et~al.}(2015{\natexlab{b}}){Ogihara}, {Morbidelli}, \&
  {Guillot}}]{ogi15b}
{Ogihara}, M., {Morbidelli}, A., \& {Guillot}, T. 2015{\natexlab{b}}, \aap,
  584, L1

\bibitem[{{Oka} {et~al.}(2011){Oka}, {Nakamoto}, \& {Ida}}]{oka11}
{Oka}, A., {Nakamoto}, T., \& {Ida}, S. 2011, \apj, 738, 141

\bibitem[{{Okuzumi} \& {Hirose}(2011)}]{oh11}
{Okuzumi}, S. \& {Hirose}, S. 2011, \apj, 742, 65

\bibitem[{{Okuzumi} {et~al.}(2014){Okuzumi}, {Takeuchi}, \& {Muto}}]{oku14}
{Okuzumi}, S., {Takeuchi}, T., \& {Muto}, T. 2014, \apj, 785, 127

\bibitem[{{Owen} {et~al.}(2011){Owen}, {Ercolano}, \& {Clarke}}]{owe11}
{Owen}, J.~E., {Ercolano}, B., \& {Clarke}, C.~J. 2011, \mnras, 412, 13

\bibitem[{{Owen} {et~al.}(2010){Owen}, {Ercolano}, {Clarke}, \&
  {Alexander}}]{owe10}
{Owen}, J.~E., {Ercolano}, B., {Clarke}, C.~J., \& {Alexander}, R.~D. 2010,
  \mnras, 401, 1415

\bibitem[{{Paardekooper} {et~al.}(2011){Paardekooper}, {Baruteau}, \&
  {Kley}}]{paa11}
{Paardekooper}, S.-J., {Baruteau}, C., \& {Kley}, W. 2011, \mnras, 410, 293

\bibitem[{{Pelletier} \& {Pudritz}(1992)}]{pp92}
{Pelletier}, G. \& {Pudritz}, R.~E. 1992, \apj, 394, 117

\bibitem[{{Pessah} {et~al.}(2006){Pessah}, {Chan}, \& {Psaltis}}]{pes06}
{Pessah}, M.~E., {Chan}, C.-K., \& {Psaltis}, D. 2006, \mnras, 372, 183

\bibitem[{{Ricci} {et~al.}(2010){Ricci}, {Testi}, {Natta}, {Neri}, {Cabrit}, \&
  {Herczeg}}]{ric10}
{Ricci}, L., {Testi}, L., {Natta}, A., {et~al.} 2010, \aap, 512, A15

\bibitem[{{Rothstein} \& {Lovelace}(2008)}]{rl08}
{Rothstein}, D.~M. \& {Lovelace}, R.~V.~E. 2008, \apj, 677, 1221

\bibitem[{{Ruden} \& {Lin}(1986)}]{rl86}
{Ruden}, S.~P. \& {Lin}, D.~N.~C. 1986, \apj, 308, 883

\bibitem[{{Sai} {et~al.}(2013){Sai}, {Katoh}, {Terada}, \& {Ono}}]{sai13}
{Sai}, K., {Katoh}, Y., {Terada}, N., \& {Ono}, T. 2013, \apj, 767, 165

\bibitem[{{Salmeron} {et~al.}(2011){Salmeron}, {K{\"o}nigl}, \&
  {Wardle}}]{sal11}
{Salmeron}, R., {K{\"o}nigl}, A., \& {Wardle}, M. 2011, \mnras, 412, 1162

\bibitem[{{Sano} {et~al.}(1998){Sano}, {Inutsuka}, \& {Miyama}}]{san98}
{Sano}, T., {Inutsuka}, S.-i., \& {Miyama}, S.~M. 1998, \apjl, 506, L57

\bibitem[{{Sano} {et~al.}(2004){Sano}, {Inutsuka}, {Turner}, \&
  {Stone}}]{san04}
{Sano}, T., {Inutsuka}, S.-i., {Turner}, N.~J., \& {Stone}, J.~M. 2004, \apj,
  605, 321

\bibitem[{{Sano} {et~al.}(2000){Sano}, {Miyama}, {Umebayashi}, \&
  {Nakano}}]{san00}
{Sano}, T., {Miyama}, S.~M., {Umebayashi}, T., \& {Nakano}, T. 2000, \apj, 543,
  486

\bibitem[{{Shakura} \& {Sunyaev}(1973)}]{ss73}
{Shakura}, N.~I. \& {Sunyaev}, R.~A. 1973, \aap, 24, 337

\bibitem[{{Shu} {et~al.}(1994){Shu}, {Najita}, {Ostriker}, {Wilkin}, {Ruden},
  \& {Lizano}}]{shu94}
{Shu}, F., {Najita}, J., {Ostriker}, E., {et~al.} 1994, \apj, 429, 781

\bibitem[{{Shu} {et~al.}(1993){Shu}, {Johnstone}, \& {Hollenbach}}]{shu93}
{Shu}, F.~H., {Johnstone}, D., \& {Hollenbach}, D. 1993, Icarus, 106, 92

\bibitem[{{Simon} {et~al.}(2013){Simon}, {Bai}, {Armitage}, {Stone}, \&
  {Beckwith}}]{sim13}
{Simon}, J.~B., {Bai}, X.-N., {Armitage}, P.~J., {Stone}, J.~M., \& {Beckwith},
  K. 2013, \apj, 775, 73

\bibitem[{{Simon} {et~al.}(2011){Simon}, {Hawley}, \& {Beckwith}}]{sim11}
{Simon}, J.~B., {Hawley}, J.~F., \& {Beckwith}, K. 2011, \apj, 730, 94

\bibitem[{{Suzuki} {et~al.}(2013){Suzuki}, {Imada}, {Kataoka}, {Kato},
  {Matsumoto}, {Miyahara}, \& {Tsuneta}}]{suz13}
{Suzuki}, T.~K., {Imada}, S., {Kataoka}, R., {et~al.} 2013, \pasj, 65, 98

\bibitem[{{Suzuki} \& {Inutsuka}(2009)}]{si09}
{Suzuki}, T.~K. \& {Inutsuka}, S.-i. 2009, \apjl, 691, L49

\bibitem[{{Suzuki} \& {Inutsuka}(2014)}]{si14}
{Suzuki}, T.~K. \& {Inutsuka}, S.-i. 2014, \apj, 784, 121

\bibitem[{{Suzuki} {et~al.}(2010){Suzuki}, {Muto}, \& {Inutsuka}}]{suz10}
{Suzuki}, T.~K., {Muto}, T., \& {Inutsuka}, S.-i. 2010, \apj, 718, 1289

\bibitem[{{Takagi} {et~al.}(2015){Takagi}, {Itoh}, {Arai}, {Sai}, \&
  {Oasa}}]{tak15}
{Takagi}, Y., {Itoh}, Y., {Arai}, A., {Sai}, S., \& {Oasa}, Y. 2015, \pasj, 67,
  87

\bibitem[{{Takagi} {et~al.}(2014){Takagi}, {Itoh}, \& {Oasa}}]{tak14}
{Takagi}, Y., {Itoh}, Y., \& {Oasa}, Y. 2014, \pasj, 66, 88

\bibitem[{{Takeuchi} \& {Okuzumi}(2014)}]{to14}
{Takeuchi}, T. \& {Okuzumi}, S. 2014, \apj, 797, 132

\bibitem[{{Tanaka} {et~al.}(2002){Tanaka}, {Takeuchi}, \& {Ward}}]{tan02}
{Tanaka}, H., {Takeuchi}, T., \& {Ward}, W.~R. 2002, \apj, 565, 1257

\bibitem[{{Tanaka} {et~al.}(2013){Tanaka}, {Nakamoto}, \& {Omukai}}]{tan13}
{Tanaka}, K.~E.~I., {Nakamoto}, T., \& {Omukai}, K. 2013, \apj, 773, 155

\bibitem[{{Turner} \& {Sano}(2008)}]{ts08}
{Turner}, N.~J. \& {Sano}, T. 2008, \apjl, 679, L131

\bibitem[{{Velikhov}(1959)}]{vel59}
{Velikhov}, E.~P. 1959, Zh. Eksp. Teor. Fiz., 36, 1398

\bibitem[{{Weidenschilling}(1977)}]{wei77}
{Weidenschilling}, S.~J. 1977, \mnras, 180, 57

\bibitem[{{Wood} {et~al.}(2014){Wood}, {M{\"u}ller}, {Redfield}, \&
  {Edelman}}]{wood14}
{Wood}, B.~E., {M{\"u}ller}, H.-R., {Redfield}, S., \& {Edelman}, E. 2014,
  \apjl, 781, L33

\bibitem[{{Wood} {et~al.}(2005){Wood}, {M{\"u}ller}, {Zank}, {Linsky}, \&
  {Redfield}}]{wood05}
{Wood}, B.~E., {M{\"u}ller}, H.-R., {Zank}, G.~P., {Linsky}, J.~L., \&
  {Redfield}, S. 2005, \apjl, 628, L143

\bibitem[{{Zanni} \& {Ferreira}(2013)}]{zf13}
{Zanni}, C. \& {Ferreira}, J. 2013, \aap, 550, A99

\end{thebibliography}

\begin{appendix}
\section{Derivation of the equation for the surface density}
\label{sec:apsig}
In this appendix (\ref{sec:apsig}), we derive Eq. (\ref{eq:sgmevl})
from the conservation equations for angular momentum and mass.
Under the axisymmetric approximation, a general MHD expression of
the conservation of angular momentum \citep[e.g.,][]{bh98} is
\begin{displaymath}
\frac{\partial}{\partial t}(\rho r v_{\phi}) + \frac{1}{r}
\frac{\partial}{\partial r}\left[r^2\left(\rho v_r v_{\phi}
  -\frac{B_r B_{\phi}}{4\pi}\right)\right]
\end{displaymath}
\begin{equation}
  + \frac{\partial}{\partial z}\left[r\left(\rho v_{\rm \phi} v_z
    -\frac{B_{\phi}B_z}{4\pi}\right)\right]  = 0.
\label{eq:angmom}
\end{equation}
The azimuthal velocity, $v_{\phi}$, is decomposed into the mean Keplerian flow and
perturbation, 
\begin{equation}
  v_{\phi} = r\Omega + \delta v_{\phi}.
  \label{eq:vphdcmp}
\end{equation}
We use the $\alpha$ prescription \citep{ss73} for the second and third terms
of Eq. (\ref{eq:angmom}): 
\begin{displaymath}
\rho v_r v_{\phi} - \frac{B_r B_{\phi}}{4\pi} = \rho v_r r\Omega
+ \rho \left(v_r \delta v_{\phi} - \frac{B_r B_{\phi}}{4\pi \rho}\right)
\end{displaymath}
\begin{equation}
  \equiv \rho v_r r\Omega + \rho \alpha_{r\phi}c_{\rm s}^2, 
  \label{eq:Mxrp}
\end{equation}
and
\begin{displaymath}
\rho v_{\phi} v_z - \frac{B_{\phi}B_z}{4\pi} = \rho r\Omega v_z
+ \rho \left(\delta v_{\phi}v_z - \frac{B_{\phi}B_z}{4\pi \rho}\right)
\end{displaymath}
\begin{equation}
  \equiv \rho r\Omega v_z + \rho \alpha_{\phi z}c_{\rm s}^2.
  \label{eq:Mxpz}
\end{equation}
We integrate Eq. (\ref{eq:angmom}) along the vertical direction, $z$,
with Eqs. (\ref{eq:Mxrp}) and (\ref{eq:Mxpz}) from the bottom surface
to the top surface of a disc, and we have 
\begin{displaymath}
\frac{\partial}{\partial t}(\Sigma r^3\Omega) + \frac{\partial}{\partial r}
\left[r^2\Sigma \left(v_r r\Omega + \overline{\alpha_{r\phi}}c_{\rm s}^2\right)\right]
\end{displaymath}
\begin{equation}
  +r^2 \left[(\rho v_z)_{\rm w}r\Omega + \overline{\alpha_{\phi z}}
    (\rho c_{\rm s}^2)_{\rm mid}\right]= 0,
  \label{eq:amintgz}
\end{equation}
where
$\overline{\alpha_{r\phi}}=\int \rho \alpha_{r\phi}dz / \Sigma$ is
the mass-weighted vertical average. 
The third term, which represents the angular momentum loss from both
surfaces, is derived from 
\begin{equation}
  \left[\rho r \Omega v_z + \rho \alpha_{\phi z}c_{\rm s}^2\right]_{\rm w}
  =[\rho v_z]_{\rm w} r\Omega + (\rho c_{\rm s}^2)_{\rm mid}
  \overline{\alpha_{\phi z}}, 
\end{equation}
where the subscript w stands for disc wind. $\overline{\alpha_{\phi z}}$
is the angular momentum loss by the $\phi z$ component of the
stress normalized by the density and the sound speed at the midplane,
Eq. (\ref{eq:phizstress}).

The equation of mass conservation is 
\begin{equation}
  \frac{\partial \Sigma}{\partial t} + \frac{1}{r}\frac{\partial}{\partial r}
  (r\Sigma v_r) + (\rho v_z)_{\rm w} =0
  \label{eq:masscont}
\end{equation}
By combining Eq. (\ref{eq:masscont}) multiplied by $r^3\Omega$ and
Eq. (\ref{eq:amintgz}), we have
\begin{equation}
  r\Sigma v_r \frac{\partial}{\partial r}(r^2 \Omega)
  + \frac{\partial}{\partial r}(r^2 \Sigma \overline{\alpha_{r\phi}}c_{\rm s}^2)
  + r^2\overline{\alpha_{\phi z}}(\rho c_{\rm s}^2)_{\rm mid} = 0,
\end{equation}
which determines the accretion rate,
\begin{equation}
  r\Sigma v_r = - \frac{2}{r\Omega} \left[\frac{\partial}{\partial r}
    (r^2 \Sigma \overline{\alpha_{r\phi}}c_{\rm s}^2)
    + r^2\overline{\alpha_{\phi z}}(\rho c_{\rm s}^2)_{\rm mid}\right], 
  \label{eq:accvel}
\end{equation}
where we here assumed the Keplerian rotation to derive
$\frac{\partial}{\partial r} (r^2\Omega) = \frac{r\Omega}{2}$.

By substituting Eq. (\ref{eq:accvel}) into Eq. (\ref{eq:masscont}),
we finally have the equation for the time evolution of $\Sigma$
(Eq. \ref{eq:sgmevl}):
\begin{displaymath}
\frac{\partial \Sigma}{\partial t} - \frac{1}{r}\frac{\partial}{\partial r}
\left[\frac{2}{r\Omega}\left\{\frac{\partial}{\partial r}(r^2 \Sigma
  \overline{\alpha_{r\phi}}c_{\rm s}^2) + r^2 \overline{\alpha_{\phi z}}
  (\rho c_{\rm s}^2)_{\rm mid} \right\}\right]
\end{displaymath}
\begin{displaymath}
+ (\rho v_z)_{\rm w} = 0, 
\end{displaymath}

\section{Energetics of accretion discs}
\label{sec:dereng}
A general MHD expression of the total energy conservation
under the axisymmetric approximation is
\begin{displaymath}
\frac{\partial}{\partial t}\left[\frac{1}{2}\rho v^2 +\rho \Phi
  + \frac{p}{\gamma-1} + \frac{B^2}{8\pi}\right] 
\end{displaymath}
\begin{displaymath}
+ \frac{1}{r}\frac{\partial}{\partial r}\left[r\left\{v_r\left(\frac{1}{2}\rho v^2
+\rho\Phi + \frac{\gamma}{\gamma -1}p + \frac{B_{\phi}^2+B_z^2}{4\pi}
\right) \right.\right.
\end{displaymath}
\begin{displaymath}
\left.\left.- \frac{B_r}{4\pi}(v_{\phi}B_{\phi}+v_z B_z)+F_{{\rm ot},r}\right\}\right]
\end{displaymath}
\begin{displaymath}
+\frac{\partial}{\partial z}\left[v_z\left(\frac{1}{2}\rho v^2
+\rho\Phi + \frac{\gamma}{\gamma -1}p + \frac{B_r^2+B_{\phi}^2}{4\pi}
\right) \right.
\end{displaymath}
\begin{equation}
  \left.- \frac{B_z}{4\pi}(v_r B_r + v_{\phi}B_{\phi}) + F_{{\rm ot},z}\right]= 0, 
\label{eq:toteng}
\end{equation}
where we refer to Eq. (\ref{eq:totengMHD}) for the definition of each variable.
Decomposing $v_{\phi}$ by Eq. (\ref{eq:vphdcmp}) and assuming $r\Omega \gg
v_r, \delta v_{\phi}, v_z, c_s, B/\sqrt{4\pi\rho}$ in a disc, we rewrite Eq.
(\ref{eq:toteng}) with leaving dominant terms. The time-derivative term becomes
\begin{displaymath}
\frac{\partial}{\partial t}\left[\frac{1}{2}\rho v^2 + \frac{p}{\gamma-1}
  +\rho \Phi + \frac{B^2}{8\pi}\right] \approx
\frac{\partial}{\partial t}\left[\frac{1}{2}\rho v^2 +\rho \Phi\right]
\end{displaymath}
\begin{equation}
  \approx  \frac{\partial}{\partial t}\left[\frac{1}{2}\rho(r\Omega
    + \delta v_{\phi})^2 - \rho r^2\Omega^2 \right] \approx
  \frac{\partial}{\partial t}\left(-\frac{1}{2}\rho r^2\Omega^2\right), 
  \label{eq:tderiv}
\end{equation}
where we set $r\Omega \delta v_{\phi}=0$ after the azimuthal average.
The $r$-derivative term, except for $F_{{\rm ot},r}$, can be approximated as
\begin{displaymath}
\frac{\partial}{\partial r}\left[r\left\{v_r\left(\frac{1}{2}\rho v^2
+\rho\Phi + \frac{\gamma}{\gamma -1}p + \frac{B_{\phi}^2+B_z^2}{4\pi}
\right) \right.\right.
\end{displaymath}
\begin{displaymath}
  \left.\left.- \frac{B_r}{4\pi}(v_{\phi}B_{\phi}+v_z B_z)\right\}\right]
\end{displaymath}
\begin{displaymath}
\approx \frac{\partial}{\partial r}\left[r\left\{v_r\left(\frac{1}{2}\rho v^2
  +\rho\Phi\right) - \frac{B_r}{4\pi}v_{\phi}B_{\phi}\right\}\right]
\end{displaymath}
\begin{displaymath}
  \approx \frac{\partial}{\partial r}\left[r\left\{-\rho v_r
    \frac{r^2\Omega^2}{2}+\rho r\Omega\left(v_r\delta v_{\phi}
    -\frac{B_r B_{\phi}}{4\pi\rho}\right)\right\}\right]
\end{displaymath}
  \begin{equation}
  =\frac{\partial}{\partial r}\left[r\left\{-\rho v_r
    \frac{r^2\Omega^2}{2}+\rho r\Omega\alpha_{r\phi}c_{\rm s}^2\right\}\right],
  \label{eq:rderiv}
\end{equation} 
where the second $\approx$ is derived from 
$v_r\left(\frac{v^2}{2} + \Phi\right)\approx  v_r\left[
  \frac{(r\Omega+\delta v_{\phi})^2}{2}
  - r^2\Omega^2\right]\approx - v_r \frac{r^2\Omega^2}{2}
+ \rho r\Omega v_r \delta v_{\phi}$, and for the last equality Eq.
(\ref{eq:Mxrp}) is used. 
We set the $z$-derivative term, except for $F_{{\rm ot}.z}$, to be 
\begin{displaymath}
\frac{\partial}{\partial z}\left[v_z\left(\frac{1}{2}\rho v^2
+\rho\Phi + \frac{\gamma}{\gamma -1}p + \frac{B_r^2+B_{\phi}^2}{4\pi}
\right) \right.
\end{displaymath}
\begin{equation}
  \left.- \frac{B_z}{4\pi}(v_r B_r + v_{\phi}B_{\phi})\right]
\equiv \frac{\partial}{\partial z}(\rho v_z E_{\rm w}).
\label{eq:zderiv}
\end{equation}
In the wind region, the kinetic energy will eventually dominate
\citep{pp92},  
\begin{equation}
  E_{\rm w} \approx \frac{v_z^2}{2} \;\; ({\rm z\Rightarrow \infty}), 
  \label{eq:Edwinfty}
\end{equation}
provided that the disc wind is accelerated with increasing $z$.  

By substituting Eqs. (\ref{eq:tderiv}) -- (\ref{eq:zderiv}) into Eq.
(\ref{eq:toteng}), we obtain
\begin{displaymath}
  \frac{\partial}{\partial t}\left(-\rho \frac{r^2\Omega^2}{2}\right) 
  +\frac{1}{r}\frac{\partial}{\partial r}\left[r\left\{-\rho v_r
    \frac{r^2\Omega^2}{2} + \rho r\Omega\alpha_{r\phi} c_{\rm s}^2 + F_{{\rm ot},r}
    \right\}\right]
\end{displaymath}
  \begin{equation}
  +\frac{\partial}{\partial z}(\rho v_z E_{\rm w} + F_{{\rm ot},z})= 0
  \label{eq:toteaprx}
\end{equation}

We integrate Eq. (\ref{eq:toteaprx}) from the bottom surface to the top
surface along $z$:
\begin{displaymath}
\frac{\partial}{\partial t}\left(-\Sigma \frac{r^2\Omega^2}{2}\right) 
+\frac{1}{r}\frac{\partial}{\partial r}\left[r\left\{-\Sigma v_r
  \frac{r^2\Omega^2}{2} + \Sigma r\Omega\overline{\alpha_{r\phi}}
  c_{\rm s}^2\right\}\right]
\end{displaymath}
\begin{equation}
  + (\rho v_z)_{\rm w} E_{\rm w}+ F_{\rm rad} =0, 
  \label{eq:totezint1}
\end{equation}
where $(\rho v_z)_{\rm w} E_{\rm w}$ and $F_{\rm rad}$ are the energy loss by disc
winds and radiation from the top and bottom surfaces. Here $F_{\rm rad}$ is from
$F_{\rm ot}$. 
By substituting Eq. (\ref{eq:accvel}) into Eq. (\ref{eq:totezint1}),
we have
\begin{displaymath}
\frac{\partial}{\partial t}\left(-\Sigma \frac{r^2\Omega^2}{2}\right) 
+\frac{1}{r}\frac{\partial}{\partial r}\left[r\Omega\left\{
\frac{\partial}{\partial r}(r^2 \Sigma
\overline{\alpha_{r\phi}}c_{\rm s}^2) + r^2 \overline{\alpha_{\phi z}}
(\rho c_{\rm s}^2)_{\rm mid} \right\}\right.
\end{displaymath}
\begin{equation}
  \left. + r^2\Omega\Sigma\overline{\alpha_{r\phi}}
  c_{\rm s}^2\right] + (\rho v_z)_{\rm w} E_{\rm w}+ F_{\rm rad} =0, 
\label{eq:totezint2}
\end{equation}

By multiplying Eq. (\ref{eq:sgmevl}) by $r^2\Omega^2/2$, we have
\begin{displaymath}
\frac{\partial}{\partial t}\left(\Sigma \frac{r^2\Omega^2}{2}\right)
-r^2\Omega^2\frac{\partial}{\partial r}
\left[\frac{1}{r\Omega}\left\{\frac{\partial}{\partial r}(r^2 \Sigma
\overline{\alpha_{r\phi}}c_{\rm s}^2) + r^2 \overline{\alpha_{\phi z}}
(\rho c_{\rm s}^2)_{\rm mid} \right\}\right]
\end{displaymath}
\begin{equation}
  + (\rho v_z)_{\rm w}\frac{r^2\Omega^2}{2} =0.
  \label{eq:sgmro2}
\end{equation}
By combining Eqs. (\ref{eq:totezint2}) and (\ref{eq:sgmro2}), we finally obtain
a simple relation for the energetics of disc wind, Eqs. (\ref{eq:engcnt1})
\& (\ref{eq:engcnt2})
\begin{eqnarray}
  & & \hspace{-1cm}
  (\rho v_z)_{\rm w}\left(E_{\rm w}+\frac{r^2\Omega^2}{2}\right) + F_{\rm rad}
  \nonumber\\
 &=&\frac{\Omega}{r}\left[\frac{\partial}{\partial r}(r^2 \Sigma
  \overline{\alpha_{r\phi}}c_{\rm s}^2) + r^2
  \overline{\alpha_{\phi z}}(\rho c_{\rm s}^2)_{\rm mid}\right]
  -\frac{1}{r}\frac{\partial}{\partial r}(r^2 \Sigma\Omega
  \overline{\alpha_{r\phi}}c_{\rm s}^2) \nonumber \\
  &=&\frac{3}{2}\Omega\Sigma\overline{\alpha_{r\phi}}c_{\rm s}^2 + r\Omega
  \overline{\alpha_{\phi z}}(\rho c_{\rm s}^2)_{\rm mid}  \nonumber
\end{eqnarray}

When the disc wind is neglected, $(\rho v_z)_{\rm w}=0$,
  $\overline{\alpha_{\phi z}}=0$, Eq. (\ref{eq:engcnt2}) is simplified to
\begin{equation}
  \sigma_{\rm SB} T^4 = \frac{3}{4}\Omega\Sigma \overline{\alpha_{r\phi}}c_{\rm s}^2
  \label{eq:visdsk}
\end{equation}
where we use Eq. (\ref{eq:SB}).
Since the mass accretion rate is approximated as $\dot{M}_r=-2\pi \Sigma r v_r
\approx 2\pi \Sigma r (\overline{\alpha_{r\phi}}c_{\rm s}^2/r\Omega)$, 
Eq. (\ref{eq:visdsk}) is rewritten as
\begin{equation}
  \sigma_{\rm SB} T^4 = \frac{3}{8\pi}\dot{M}_r \Omega^2 = \frac{3}{8\pi}
  \frac{GM_{\star} \dot{M}_r}{r^3}, 
\end{equation}
which is consistent with the expression for the standard accretion disc in
the outer region \citep{ss73}.

\end{appendix}

\end{document}